\documentclass[aps,onecolumn,nofootinbib,tightenlines,superscriptaddress]{revtex4}

\usepackage{amssymb,amstext,amsmath,amsthm,slashed}
\usepackage[dvips]{graphicx}
\usepackage{latexsym}
\usepackage{psfrag}
\usepackage{amsfonts}
\usepackage{bbm} 
\usepackage{color}
\usepackage{verbatim}
\usepackage{bbm}
\usepackage{dcolumn}
\usepackage{tabularx}
\usepackage{leftidx}
\usepackage{tensor}
\usepackage{boxedminipage}
\usepackage{fancyvrb}
\usepackage{float}
\usepackage{wrapfig}
\usepackage{booktabs}
\usepackage{dsfont} 
\usepackage{mathrsfs}
\usepackage[hidelinks]{hyperref}

\newtheorem{Def}{Definition}[section]

\newtheorem{Prp}[Def]{Proposition}
\newtheorem{Lemma}[Def]{Lemma}

\newcommand{\Proof}{\begin{proof}}
\newcommand{\QED}{\end{proof} \noindent}

\newcommand{\1}{\mbox{\rm 1 \hspace{-1.05 em} 1}}

\begin{document}

\title{The Massive Dirac Equation in the Kerr Geometry: Separability \\ in Eddington--Finkelstein-type Coordinates and Asymptotics}

\vskip.9in

\author{Christian R\"oken\vspace{0.1cm}}

\vskip.9in

\affiliation{Universit\"at Regensburg, Fakult\"at f\"ur Mathematik, 93040 Regensburg, Germany \footnotetext{e-mail: christian.roeken@mathematik.uni-regensburg.de} \vspace{0.1cm}}

\affiliation{Departamento de Geometr\'ia y Topolog\'ia, Facultad de Ciencias - Universidad de Granada, Campus de Fuentenueva s/n, 18071 Granada, Spain \vspace{0.4cm}}

\vskip.9in 

\date{February 2019}

\begin{abstract}
\vspace{0.4cm} \noindent \textbf{\footnotesize ABSTRACT.} \, The separability of the massive Dirac equation in the non-extreme Kerr geometry in horizon-penetrating advanced Eddington--Finkelstein-type coordinates is shown. To this end, the Kerr geometry is described in the Newman--Penrose formalism by a regular Carter tetrad and the Dirac spinors and matrices are defined in a chiral Newman--Penrose dyad representation. Applying Chandrasekhar's mode ansatz, the Dirac equation is separated into radial and angular systems of ordinary differential equations. Asymptotic radial solutions at infinity, the event horizon, and the Cauchy horizon are derived, and the decay of the associated errors is analyzed. Moreover, specific aspects of the angular eigenfunctions and eigenvalues are discussed. Finally, as an application, the scattering of massive Dirac particles by the gravitational field of a rotating Kerr black hole is studied. This work provides the basis for a Hamiltonian formulation of the massive Dirac equation in the non-extreme Kerr geometry in horizon-penetrating coordinates and for the construction of an integral spectral representation of the Dirac propagator that yields the dynamics of Dirac particles outside, across, and inside the event horizon, up to the Cauchy horizon. 
\end{abstract}

\setcounter{tocdepth}{2}

\vspace{0.1cm}

\maketitle

\tableofcontents

%-----------------------------------------------------
\section{Introduction}
%-----------------------------------------------------

\noindent Over the last five decades, the dynamics of relativistic spin-$1/2$ fermions (Dirac particles) in the Kerr geometry was studied extensively employing different approaches. The probably most established approach is Chandrasekhar's mode analysis \cite{Chandra, ChaDet, ChandraBook}, where the massive Dirac equation is separated by means of time and azimuthal angle modes and a specific product ansatz for the radial and polar angle dependencies in the Dirac spinors, which results in radial wave equations and a first-order system of angular ordinary differential equations (ODEs). Within this framework, many physical processes like the emission and absorption of Dirac particles by rotating Kerr black holes, Kerr black hole stability under fermionic field perturbations, and superradiance were investigated \cite{BriWhee, Page, PrTe, Teuk, Unruh}. However, it does not give rise to a description of the full time-dependent dynamics of Dirac particles in the Kerr geometry. This was first analyzed in the framework of scattering theory in \cite{DH, Bat, ChaMuk, FHM, Daude, HaNic, Nic}. More recently, a somewhat different approach to define the dynamics of Dirac particles using an integral spectral representation of the Dirac propagator in the Hamiltonian framework was developed \cite{FKSM1, FKSM2, FKSM3}. This method combines results from both functional analysis and Chandrasekhar's mode analysis, making it possible to derive sharp decay rates for Dirac spinors as well as estimates for the probability of a Dirac particle to fall into a Kerr black hole or escape to infinity.

The basis of the mode analysis approach is Chandrasekhar's famous discovery that the massive Dirac equation in the non-extreme Kerr geometry expressed in Boyer--Lindquist coordinates is separable, which was worked out in his original article \cite{Chandra} from 1976 and led to a major breakthrough in the field of black hole physics. At that time, this remarkable result came a bit as a surprise because the Dirac equation was not expected to be separable in the Kerr geometry. Despite the tremendous impact of this discovery, the validity of the associated solutions is naturally restricted to those regions of the Kerr geometry where the Boyer--Lindquist coordinates are well-defined. And as they become singular at the event and the Cauchy horizon, respectively, the dynamics of Dirac particles near and across these horizons cannot be properly described in these coordinates. In this article, we resolve this problem by employing a particular analytic extension of the Boyer--Lindquist coordinates that is regular at the horizons, thus, covering both the exterior and interior black hole regions and allowing for well-defined transitions of Dirac particles across the horizons. This analytic extension, the so-called advanced Eddington--Finkelstein-type coordinates, also features a proper time function (unlike the original advanced Eddington--Finkelstein null coordinates \cite{Edd, Fink}), which is required for -- and gives rise to -- the Hamiltonian formulation of the Dirac equation as the fundamental framework for the construction of the integral spectral representation of the Dirac propagator. We point out that since the transformation from Boyer--Lindquist coordinates to advanced Eddington--Finkelstein-type coordinates is singular at the event and the Cauchy horizon, and hence non-trivial, a careful mathematical analysis on this issue is essential. Furthermore, as the mixing of the time and the azimuthal angle variable that arises in this transformation leads to a symmetry breaking of structures inherent to Boyer--Lindquist coordinates, it is a priori not clear that the separation of variables property of the Dirac equation is conserved in advanced Eddington--Finkelstein-type coordinates. We also note that in the recent article \cite{Dolan3}, a similar coordinate system is employed. This work concentrates on certain physical aspects of the massive Dirac equation in horizon-penetrating coordinates, namely, using numerical methods, the energy spectrum and the decay rates of bound states are computed. Here, on the other hand, the focus lies on rather mathematical aspects, i.e., we derive the Newman--Penrose formalism of the Dirac equation in horizon-penetrating coordinates, present a detailed analysis of the asymptotics of the Dirac spinors at space-like infinity, the event horizon, and the Cauchy horizon, which includes accurate estimates of the decay of the associated errors, and discuss the nature and specific properties of the angular eigenfunctions and eigenvalues. Besides, instead of using advanced Eddington--Finkelstein-type coordinates, one may as well choose different horizon-penetrating coordinates such as Doran coordinates \cite{Doran} or generalized Kruskal--Szekeres coordinates \cite{Kruskal, ChandraBook, CR}. These are, however, far more complicated in their handling and result in more intricate computations.

We perform the mode analysis of the massive Dirac equation in the non-extreme Kerr geometry in horizon-penetrating coordinates as follows. Firstly, in Section \ref{Sec2}, we describe the Kerr geometry in the Newman--Penrose formalism by a regular Carter tetrad in advanced Eddington--Finkelstein-type coordinates and calculate the associated spin coefficients. Secondly, with these quantities, we define the massive Dirac equation in the chiral representation in $2$-spinor form, for which we employ a Newman--Penrose dyad basis for the $2$-spinor space, in Section \ref{Sec3}. General overviews of the Newman--Penrose formalism and of the general relativistic Dirac equation are given in Appendices \ref{AA} and \ref{AB}. We note that the Newman--Penrose formalism is well suited for the analysis of radiative and particle transport in curved spacetimes, especially for the propagation of Dirac particles in the Kerr geometry, because it can be arranged to reflect symmetries -- or be adapted to certain aspects -- of the underlying spacetime, which results in the reduction of the number of conditional equations and leads to simplified expressions for geometric quantities. Next, considering a factorization of the Dirac spinors in time and azimuthal angle modes, we show the separability of the massive Dirac equation in horizon-penetrating advanced Eddington--Finkelstein-type coordinates into radial and angular ODE systems by means of Chandrasekhar's product ansatz. Thirdly, we determine the asymptotic radial solutions at infinity, the event horizon, and the Cauchy horizon in Sections \ref{sec3b}-\ref{sec:4D}. Moreover, we demonstrate that the associated errors have suitable decay. In Section \ref{sec:4E}, we first show that by decoupling the angular system, one obtains the usual Chandrasekhar--Page equation. Then, we discuss the nature and certain properties of both the corresponding set of eigenfunctions and the eigenvalue spectrum. Finally, in Section \ref{Sec4}, we apply the radial asymptotics at infinity and at the event horizon, as well as a normalization condition for the angular eigenfunctions, to the physical problem of scattering of massive Dirac particles by the gravitational field of a rotating Kerr black hole. More precisely, we derive a conservation law for the net current of Dirac particles in horizon-penetrating coordinates, which is in accordance with the one for Boyer--Lindquist coordinates, and further evaluate the net current at space-like infinity and at the event horizon, yielding a condition for the reflection and transmission coefficients. We find that, on the one hand, superradiance does not occur and, on the other hand, the conserved net current stays positive across the event horizon. This article provides the basis for a Hamiltonian formulation of the massive Dirac equation in the non-extreme Kerr geometry in horizon-penetrating coordinates. Within this framework, an integral spectral representation of the Dirac propagator that describes the dynamics of relativistic spin-$1/2$ fermions outside, across, and inside the event horizon, up to the Cauchy horizon, can be constructed. This is worked out in detail in \cite{RF}.

%------------------------------------------------------------------------------------------------------------------------
\section{Carter Tetrad and Spin Coefficients for the Kerr Geometry in Horizon-penetrating Advanced Eddington--Finkelstein-type Coordinates} \label{Sec2}
%------------------------------------------------------------------------------------------------------------------------

\begin{figure}[p] 
\begin{center}
\vspace{1.0cm}
\includegraphics[scale=0.70]{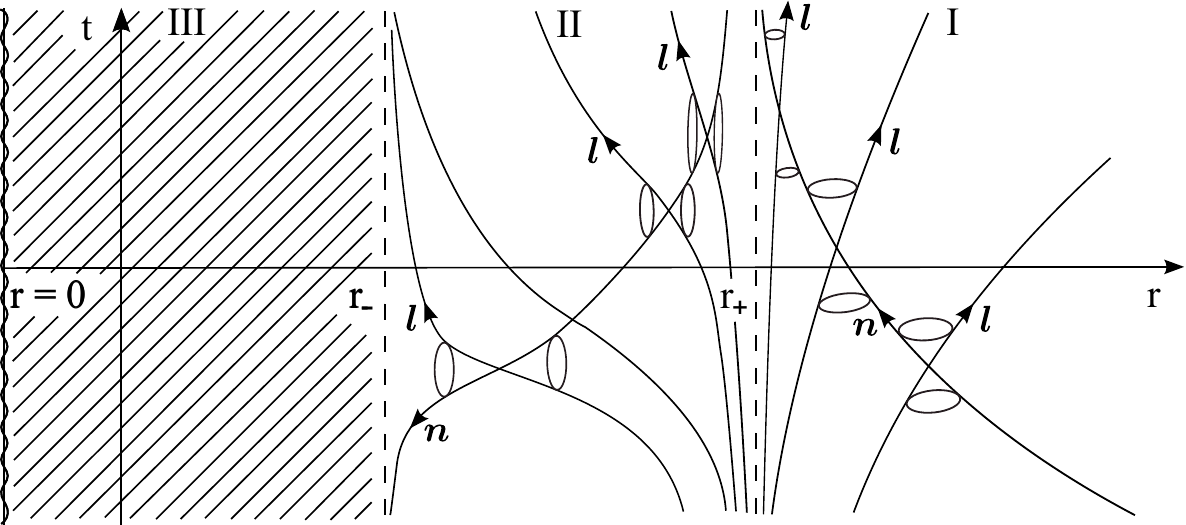}  
\caption{\small{\label{PIC1} Causal structure of the non-extreme Kerr geometry in Boyer--Lindquist coordinates. A projection onto the $(t, r)$-plane, where every point is a $2$-sphere, is presented. The real-valued Newman--Penrose null vectors $\boldsymbol{l}$ and $\boldsymbol{n}$, which point along the principal null directions, form the light cones. The light cone of an observer approaching the event horizon from outside the black hole ($r \searrow r_+$) closes up and becomes degenerate. In contrast, it opens up when the observer approaches the event horizon from inside the black hole ($r \nearrow r_+$). This stems from the fact that the roles of space and time are reversed in the black hole interior region II. When $r \rightarrow \infty$, the light cone becomes a $45^{\circ}$-Minkowski light cone because the spacetime is asymptotically flat. We note that all figures are restricted to regions I and II in order to avoid the issues that arise when one considers the ring singularity at $(r = 0, \theta = \pi/2)$ and the maximum analytic extension.}}
\end{center}
\vspace{2.0cm}
\begin{center}
\includegraphics[scale=0.70]{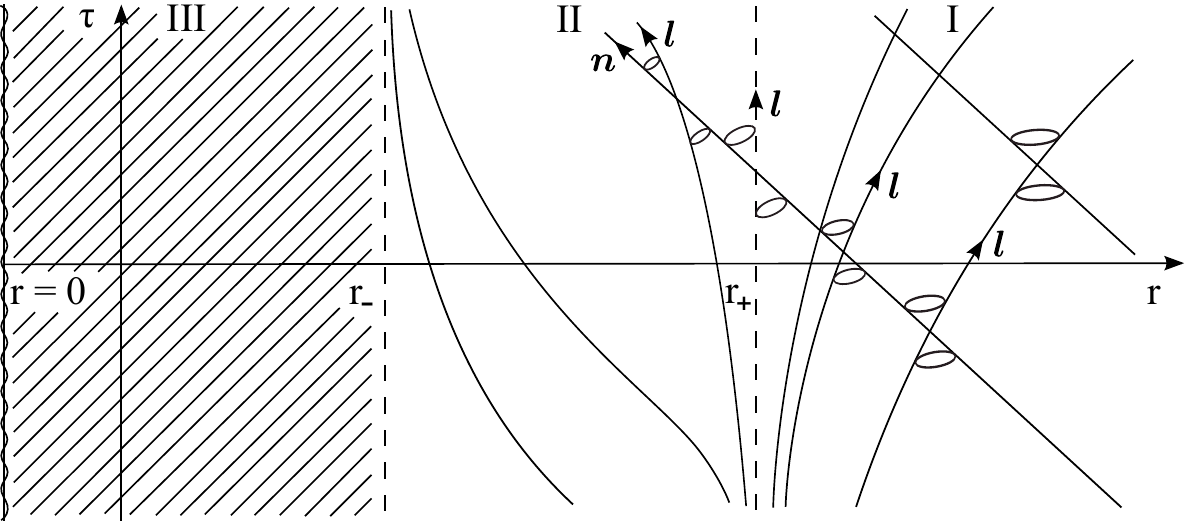} 
\caption{\small{\label{PIC2} Causal structure of the non-extreme Kerr geometry in advanced Eddington--Finkelstein-type coordinates. A projection onto the $(\tau, r)$-plane is presented. Ingoing light rays are straight lines pointing in the $\boldsymbol{n}$-direction. The light cone of an observer moving toward the event horizon from outside the black hole turns until -- after having crossed the event horizon -- its future light cone is completely in the black hole interior. This shows the trapping characteristic of event horizons.}}
\end{center}
\end{figure}

\noindent The non-extreme Kerr geometry \cite{Kerr} is a connected, orientable and time-orientable, smooth, asymptotically flat Lorentzian 4-manifold $(\mathfrak{M}, \boldsymbol{g})$ with topology $S^2 \times \mathbb{R}^2$ that consists of a differentiable manifold $\mathfrak{M}$ and a stationary, axisymmetric Lorentzian metric $\boldsymbol{g}$ with signature $(1, 3)$, which in Boyer--Lindquist coordinates $(t, r, \theta, \varphi)$ \cite{BL}, where $t \in \mathbb{R}, r \in \mathbb{R}_{> 0}, \theta \in (0, \pi)$, and $\varphi \in [0, 2 \pi)$, is given by 
\begin{equation}\label{KMBLC}
\begin{split}
\boldsymbol{g} & = \frac{\Delta}{\Sigma} \, \bigl(\textnormal{d}t - a \sin^2{(\theta)} \, \textnormal{d}\varphi\bigr) \otimes \bigl(\textnormal{d}t - a \sin^2{(\theta)} \, \textnormal{d}\varphi\bigr) - \frac{\sin^2{(\theta)}}{\Sigma} \, \bigl([r^2 + a^2] \, \textnormal{d}\varphi - a \, \textnormal{d}t\bigr) \otimes \bigl([r^2 + a^2] \, \textnormal{d}\varphi - a \, \textnormal{d}t\bigr) \\ 
& \hspace{0.4cm} - \frac{\Sigma}{\Delta} \, \textnormal{d}r \otimes \textnormal{d}r - \Sigma \, \textnormal{d}\theta \otimes \textnormal{d}\theta \, .
\end{split}
\end{equation}
The horizon function is defined by $\Delta = \Delta(r) := (r - r_+) (r - r_-) = r^2 - 2 M r + a^2$, $r_{\pm} := M \pm \sqrt{M^2 - a^2}$ denote the event and the Cauchy horizon, respectively, $M$ is the mass and $a M$ the angular momentum of the black hole with $0 \leq a < M$, and $\Sigma = \Sigma(r, \theta) := r^2 + a^2 \cos^2{(\theta)}$. We describe the Kerr geometry in terms of a local Newman--Penrose null tetrad frame that is adapted to the principal null geodesics, i.e., the tetrad coincides with the two principal null directions of the Weyl tensor. In this so-called Kinnersley frame \cite{Kinn}, since the Kerr geometry is algebraically special and of Petrov type D, we are presented with the computational advantage that the four spin coefficients $\kappa, \sigma, \lambda$, and $\nu$ vanish and only one Weyl scalar, namely $\Psi_2$, is non-zero. Accordingly, the congruences formed by the two principal null directions must be geodesic and shear-free \cite{BON}. We construct the Kinnersley tetrad directly from the tangent vectors of the principal null geodesics \cite{ChandraBook}
\begin{equation} \label{TV}
\frac{\textnormal{d}t}{\textnormal{d}\chi} = \frac{r^2 + a^2}{\Delta} \, E \, , \quad \frac{\textnormal{d}r}{\textnormal{d}\chi} = \pm E \, , \quad \frac{\textnormal{d}\theta}{\textnormal{d}\chi} = 0 \, , \quad \textnormal{and} \quad \frac{\textnormal{d}\varphi}{\textnormal{d}\chi} = \frac{a}{\Delta} \, E \, ,
\end{equation}
where $\chi$ is an affine parameter and $E$ denotes a constant, by aligning the real-valued Newman--Penrose vectors $\boldsymbol{l}$ and $\boldsymbol{n}$ with the associated principal null directions and further by choosing complex-conjugate Newman--Penrose vectors $\boldsymbol{m}$ and $\boldsymbol{\overline{m}}$ in such a way that they satisfy the conditions (\ref{NPNC})-(\ref{NPCNC}). Thus, we obtain the frame
\begin{equation}\label{KT}
\begin{split}
\boldsymbol{l} & = \frac{1}{|\Delta|} \bigl([r^2 + a^2] \, \partial_t + \Delta \, \partial_r + a \, \partial_{\varphi}\bigr) \\ \\
\boldsymbol{n} & = \frac{\textnormal{sign}(\Delta)}{2 \, \Sigma} \bigl([r^2 + a^2] \, \partial_t - \Delta \, \partial_r + a \, \partial_{\varphi}\bigr) \\ \\
\boldsymbol{m} & = \frac{1}{\sqrt{2} \bigl(r + \textnormal{i} a \cos{(\theta)}\bigr)} \bigl(\textnormal{i} a \sin{(\theta)} \, \partial_t + \partial_{\theta} + \textnormal{i} \csc{(\theta)} \, \partial_{\varphi}\bigr) \\ \\
\boldsymbol{\overline{m}} & = - \frac{1}{\sqrt{2} \bigl(r - \textnormal{i} a \cos{(\theta)}\bigr)} \bigl(\textnormal{i} a \sin{(\theta)} \, \partial_t - \partial_{\theta} + \textnormal{i} \csc{(\theta)} \, \partial_{\varphi}\bigr) 
\end{split}
\end{equation}
with the signum function
\begin{equation*}
\textnormal{sign}(\Delta) := \begin{cases} +1  & \textnormal{for} \,\,\, \Delta \geq 0 \\ -1 & \textnormal{for} \,\,\, \Delta < 0 \, . \end{cases}
\end{equation*}
We remark that the use of the absolute value of the horizon function in $\boldsymbol{l}$ and the signum function in $\boldsymbol{n}$ leads to a combined representation of the frame for the regions outside the event horizon, inside the event horizon up to the Cauchy horizon, and inside the Cauchy horizon. For the calculation of the corresponding spin coefficients, i.e., for solving the torsion-free Maurer--Cartan equation of structure (\ref{NPMCE}), one requires the dual co-tetrad of (\ref{KT}) 
\begin{equation*}
\begin{split}
\boldsymbol{l}_{\textnormal{D}} & = \textnormal{sign}(\Delta) \biggl(\textnormal{d}t - \frac{\Sigma}{\Delta} \, \textnormal{d}r - a \sin^2{(\theta)} \,  \textnormal{d}\varphi\biggr) \\ \\
\boldsymbol{n}_{\textnormal{D}} & = \frac{|\Delta|}{2 \, \Sigma} \biggl(\textnormal{d}t + \frac{\Sigma}{\Delta} \, \textnormal{d}r - a \sin^2{(\theta)} \, \textnormal{d}\varphi\biggr) \\ \\
\boldsymbol{m}_{\textnormal{D}} & = \frac{1}{\sqrt{2} \bigl(r + \textnormal{i} a \cos{(\theta)}\bigr)} \bigl(\textnormal{i} a \sin{(\theta)} \, \textnormal{d}t - \Sigma \, \textnormal{d}\theta - \textnormal{i} \, [r^2 + a^2] \sin{(\theta)} \, \textnormal{d}\varphi\bigr) \\ \\
\boldsymbol{\overline{m}}_{\textnormal{D}} & = - \frac{1}{\sqrt{2} \bigl(r - \textnormal{i} a \cos{(\theta)}\bigr)} \bigl(\textnormal{i} a \sin{(\theta)} \, \textnormal{d}t + \Sigma \, \textnormal{d}\theta - \textnormal{i} \, [r^2 + a^2] \sin{(\theta)} \, \textnormal{d}\varphi\bigr) \, .
\end{split}
\end{equation*}
However, before determining the spin coefficients, we apply a class III local Lorentz transformation (\ref{CLIIILLTT}) with parameters of the form
\begin{equation*}
\varsigma = \sqrt{\frac{|\Delta|}{2 \, \Sigma}} \quad \textnormal{and} \quad e^{\textnormal{i} \psi} = \frac{\sqrt{\Sigma}}{r - \textnormal{i} a \cos{(\theta)}} 
\end{equation*}
to the Kinnersley tetrad (\ref{KT}) in order to obtain the so-called Carter tetrad 
\begin{equation}\label{CT}
\begin{split}
\boldsymbol{l}' & = \frac{1}{\sqrt{2 \, \Sigma \, |\Delta|}} \bigl([r^2 + a^2] \, \partial_t + \Delta \, \partial_r + a \, \partial_{\varphi}\bigr) \\ \\
\boldsymbol{n}' & = \frac{\textnormal{sign}(\Delta)}{\sqrt{2 \, \Sigma \, |\Delta|}} \bigl([r^2 + a^2] \, \partial_t - \Delta \, \partial_r + a \, \partial_{\varphi}\bigr) \\ \\
\boldsymbol{m}' & = \frac{1}{\sqrt{2 \, \Sigma}} \bigl(\textnormal{i} a \sin{(\theta)} \, \partial_t + \partial_{\theta} + \textnormal{i} \csc{(\theta)} \, \partial_{\varphi}\bigr) \\ \\
\boldsymbol{\overline{m}}\hspace{0.03cm}' & = - \frac{1}{\sqrt{2 \, \Sigma}} \bigl(\textnormal{i} a \sin{(\theta)} \, \partial_t - \partial_{\theta} + \textnormal{i} \csc{(\theta)} \, \partial_{\varphi}\bigr) \, ,
\end{split}
\end{equation}
for which the dual co-tetrad reads
\begin{equation}\label{DCT}
\begin{split}
\boldsymbol{l}'_{\textnormal{D}} & = \sqrt{\frac{|\Delta|}{2 \, \Sigma}} \,\, \textnormal{sign}(\Delta) \biggl(\textnormal{d}t - \frac{\Sigma}{\Delta} \, \textnormal{d}r - a \sin^2{(\theta)} \, \textnormal{d}\varphi\biggr) \\ \\
\boldsymbol{n}'_{\textnormal{D}} & = \sqrt{\frac{|\Delta|}{2 \, \Sigma}} \biggl(\textnormal{d}t + \frac{\Sigma}{\Delta} \, \textnormal{d}r - a \sin^2{(\theta)} \, \textnormal{d}\varphi\biggr) \\ \\
\boldsymbol{m}'_{\textnormal{D}} & = \frac{1}{\sqrt{2 \, \Sigma}} \bigl(\textnormal{i} a \sin{(\theta)} \, \textnormal{d}t - \Sigma \, \textnormal{d}\theta - \textnormal{i} \, [r^2 + a^2] \sin{(\theta)} \, \textnormal{d}\varphi\bigr) \\ \\
\boldsymbol{\overline{m}}\hspace{0.03cm}'_{\textnormal{D}} & = - \frac{1}{\sqrt{2 \, \Sigma}} \bigl(\textnormal{i} a \sin{(\theta)} \, \textnormal{d}t + \Sigma \, \textnormal{d}\theta - \textnormal{i} \, [r^2 + a^2] \sin{(\theta)} \, \textnormal{d}\varphi\bigr) \, .
\end{split}
\end{equation}
This frame transforms under the composition of the discrete time and azimuthal angle reversal isometries $t \mapsto -t$ and $\varphi \mapsto - \varphi$ as 
\begin{equation*}
\boldsymbol{l}' \mapsto - \textnormal{sign}(\Delta) \, \boldsymbol{n}' \, , \,\,\,\,\,\, \boldsymbol{n}' \mapsto - \textnormal{sign}(\Delta) \, \boldsymbol{l}' \, , \,\,\,\,\,\, \boldsymbol{m}' \mapsto \boldsymbol{\overline{m}}\hspace{0.03cm}' \, , \,\,\,\,\,\, \boldsymbol{\overline{m}}\hspace{0.03cm}' \mapsto \boldsymbol{m}' \, ,
\end{equation*}
thus giving rise to only six independent spin coefficients 
\begin{equation*}
\kappa' = - \nu' \, , \quad \pi' = - \tau' \, , \quad \alpha' = - \beta' \, , \quad \sigma'  = \textnormal{sign}(\Delta) \, \lambda' \, , \quad \mu' = \textnormal{sign}(\Delta) \, \varrho' \, , \quad \epsilon' = \textnormal{sign}(\Delta) \, \gamma' \, . 
\end{equation*}
Moreover, we require a suitable coordinate system to analyze the propagation of Dirac particles across the event and the Cauchy horizon. Using Boyer--Lindquist coordinates, we have to deal with singularities at the horizons and with a reversal of the roles of space and time in between. These undesirable properties are depicted in the Finkelstein diagram in Figure \ref{PIC1}, which shows the light cone of a system that approaches -- and crosses -- the event horizon from outside the black hole. More precisely, approaching the event horizon, the light cone closes up and eventually becomes degenerate. Beyond the event horizon, it tilts over, yielding a reversal of the specific characteristics of the time and the radial variable. These issues preclude a proper study of the propagation of Dirac particles across the horizons. Hence, instead of Boyer--Lindquist coordinates, we employ advanced Eddington--Finkelstein-type coordinates, which are regular in the entire black hole spacetime except for the ring singularity at $(r = 0, \theta = \pi/2)$, and therefore allow for well-defined transitions of Dirac particles across the horizons. Furthermore, they feature a time function required for the Hamiltonian formulation of the Dirac equation and the corresponding Cauchy problem. In the following, we construct the advanced Eddington--Finkelstein-type coordinates from the tangent vectors (\ref{TV}) associated with the ingoing principal null geodesics, which are defined in terms of the usual Boyer--Lindquist coordinates. Combining the first and the second as well as the second and the fourth tangent vector, we obtain the two relations between the time and the radial coordinate
\begin{align}
\frac{\textnormal{d}t}{\textnormal{d}r} = \pm \frac{r^2 + a^2}{\Delta} \quad & \Leftrightarrow \quad t = \pm \int \frac{r^2 + a^2}{\Delta} \, \textnormal{d}r + c_{\pm} = \pm r_{\star} + c_{\pm} \label{trrel} 
\end{align}
and the two relations between the azimuthal angle and the radial coordinate 
\begin{align}
\frac{\textnormal{d}\varphi}{\textnormal{d}r} = \pm \frac{a}{\Delta} \quad & \Leftrightarrow \quad \varphi = \pm \int \frac{a}{\Delta} \, \textnormal{d}r + c'_{\pm} = \pm \frac{a}{r_+ - r_-} \ln{\biggl|\frac{r - r_+}{r - r_-}\biggr|} + c'_{\pm} \, , \label{rphirel}
\end{align}
respectively, where
\begin{equation*}
r_{\star} := r + \frac{r_+^2 + a^2}{r_+ - r_-} \ln{|r - r_+|} - \frac{r_-^2 + a^2}{r_+ - r_-} \ln{|r - r_-|}   
\end{equation*}
is the Regge--Wheeler coordinate and $c_{\pm}, c'_{\pm}$ are constants of integration. The relations for the ingoing principal null geodesics, that is the branches of (\ref{trrel}) and (\ref{rphirel}) with negative sign, constitute the transformation from Boyer--Lindquist coordinates to advanced Eddington--Finkelstein-type coordinates
\begin{equation*} 
\mathbb{R} \times \mathbb{R}_{> 0} \times (0, \pi) \times [0, 2 \pi) \rightarrow \mathbb{R} \times \mathbb{R}_{> 0} \times (0, \pi) \times [0, 2 \pi) \, , \,\,\,\,\,\,\,\, (t, r, \theta, \varphi) \mapsto (\tau, r, \theta, \phi)
\end{equation*}
with
\begin{align} 
\tau & := t + r_{\star} - r = t + \frac{r_+^2 + a^2}{r_+ - r_-} \ln{|r - r_+|} - \frac{r_-^2 + a^2}{r_+ - r_-} \ln{|r - r_-|} \label{AEFTCT} \\ \nonumber \\
\phi & := \varphi + \frac{a}{r_+ - r_-} \ln{\biggl|\frac{r - r_+}{r - r_-}\biggr|} \, . \label{AEFTCA}
\end{align}
We point out that since the relations (\ref{trrel}) and (\ref{rphirel}) only hold along the principal null geodesics, the advanced Eddington--Finkelstein-type coordinates are therefore adapted to the ingoing principal null geodesics. Moreover, in the definition of the time coordinate (\ref{AEFTCT}), the additional term $- r$ was introduced in order for $\tau$ to be a proper time function, which is explicitly verified below. These coordinates are free of singularities at the horizons and their spatio-temporal characteristics across the horizons are conserved. This can be seen in the Finkelstein diagram presented in Figure \ref{PIC2}, where we again show a system approaching -- and crossing -- the event horizon from outside the black hole, but now the light cone is non-degenerate at the event horizon and is not tilted over in between the event and the Cauchy horizon. Besides, ingoing light rays are represented simply by straight lines, as can be directly inferred from the first of the two relations between the advanced Eddington--Finkelstein-type time and radial coordinates for ingoing and outgoing principal null geodesics
%
%----------------------------------------------------------------
%
%
%----------------------------------------------------------------
%
\begin{figure}[t] 
\begin{center}
%----------------------------
\begin{minipage}{0.4\textwidth}
\includegraphics[width=0.95\textwidth]{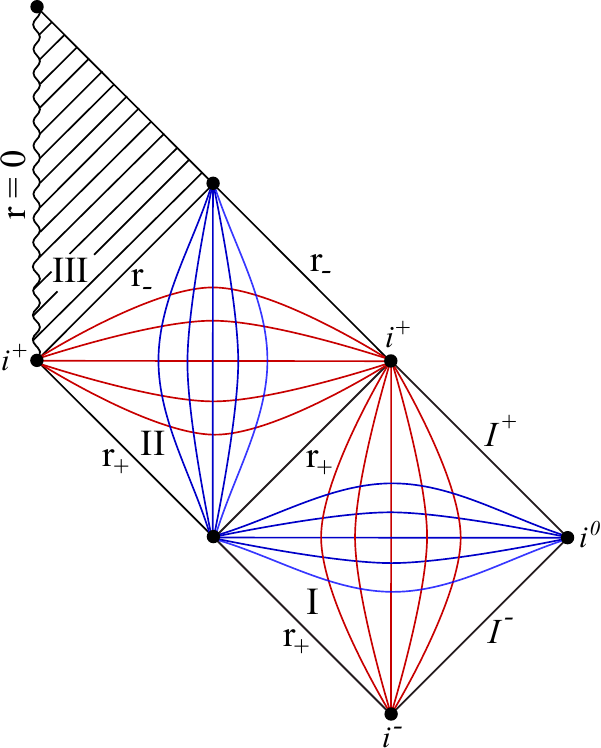} 
\end{minipage}
%----------------------------
\hspace{1.2cm}
%----------------------------
\begin{minipage}{0.4\textwidth}
\includegraphics[width=0.95\textwidth]{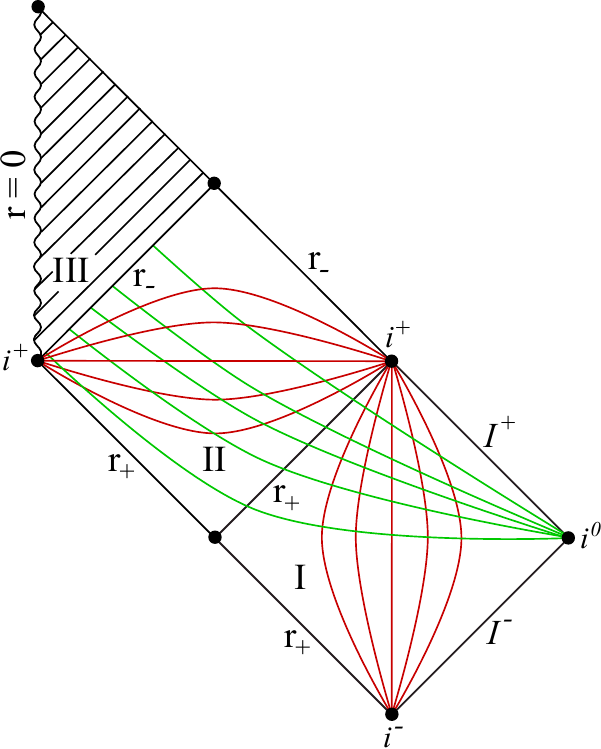} 
\end{minipage}  
%----------------------------
\caption{\small{\label{PIC3} Carter--Penrose diagrams for the non-extreme Kerr geometry in Boyer--Lindquist coordinates (left) and advanced Eddington--Finkelstein-type coordinates (right). The blue lines represent constant-$t$ hypersurfaces, the red lines constant-$r$ hypersurfaces, and the green lines constant-$\tau$ hypersurfaces. The constant-$t$ and constant-$r$ hypersurfaces are restricted to either the exterior or the interior of the black hole. Their nature changes across the event horizon, i.e., space-like hypersurfaces become time-like and vice versa. The constant-$\tau$ hypersurfaces (cut-off at the Cauchy horizon) are space-like outside and inside the black hole and smooth across the event horizon.}}
\end{center}
\end{figure}
%
%----------------------------------------------------------------
%
\begin{equation*}
\frac{\textnormal{d}\tau}{\textnormal{d}r}_{| \textnormal{in}} = - 1 \quad \textnormal{and} \quad \frac{\textnormal{d}\tau}{\textnormal{d}r}_{| \textnormal{out}} = 1 + \frac{4 M r}{\Delta} \, ,
\end{equation*}
which may be derived by inserting (\ref{AEFTCT}) into (\ref{trrel}). The metric (\ref{KMBLC}) represented in advanced Eddington--Finkelstein-type coordinates becomes
\begin{equation*} 
\begin{split}
\boldsymbol{g} & = \biggl(1 - \frac{2 M r}{\Sigma}\biggr) \textnormal{d}\tau \otimes \textnormal{d}\tau - \frac{2 M r}{\Sigma} \bigl([\textnormal{d}r - a \sin^2{(\theta)} \, \textnormal{d}\phi] \otimes \textnormal{d}\tau 
+ \textnormal{d}\tau \otimes [\textnormal{d}r - a \sin^2{(\theta)} \, \textnormal{d}\phi]\bigr) \\ 
& \hspace{0.4cm} - \biggl(1 + \frac{2 M r}{\Sigma}\biggr) \bigl(\textnormal{d}r - a \sin^2{(\theta)} \, \textnormal{d}\phi\bigr) \otimes \bigl(\textnormal{d}r - a \sin^2{(\theta)} \, \textnormal{d}\phi\bigr) - \Sigma \, \textnormal{d}\theta \otimes \textnormal{d}\theta - \Sigma \sin^2{(\theta)} \, \textnormal{d}\phi \otimes \textnormal{d}\phi \, .
\end{split}
\end{equation*}
As the associated induced metric tensor on constant-$\tau$ hypersurfaces 
\begin{equation*} 
\mathscr{G}:= (g_{\mu \nu})_{| \tau = \textnormal{const.}} = \left(\begin{array}{ccc}
\displaystyle - \biggl(1 + \frac{2 M r}{\Sigma}\biggr) & 0 & \displaystyle a \sin^2{(\theta)} \biggl(1 + \frac{2 M r}{\Sigma}\biggr) \\ \\
0 & - \Sigma & 0 \\ \\
\displaystyle a \sin^2{(\theta)} \biggl(1 + \frac{2 M r}{\Sigma}\biggr) & 0 & \displaystyle - \sin^2{(\theta)} \biggl[\Sigma + a^2 \sin^2{(\theta)} \biggl(1 + \frac{2 M r}{\Sigma}\biggr)\biggr]
\end{array}\right) ,
\end{equation*}
with the leading principal minors
\begin{equation*}
\begin{split}
\textnormal{det}(\mathscr{G}_1) & = \mathscr{G}_{1 1} = - \Sigma^{- 1} (\Sigma + 2 M r) < 0 \\
\textnormal{det}(\mathscr{G}_2) & = \mathscr{G}_{1 1} \mathscr{G}_{2 2} - \mathscr{G}_{1 2} \mathscr{G}_{2 1} = \Sigma + 2 M r > 0 \\
\textnormal{det}(\mathscr{G}_3) & = \textnormal{det}(\mathscr{G}) = - \Sigma \, (\Sigma + 2 M r) \sin^2{(\theta)} < 0 \, ,
\end{split}
\end{equation*}
is negative definite and thus Riemannian, we recognize that the coordinate $\tau$ is a proper time function. In the Carter--Penrose diagrams shown in Figure \ref{PIC3}, we depict the constant-$t$ and constant-$r$ hypersurfaces in Boyer--Lindquist coordinates (left diagram) and the constant-$\tau$ and constant-$r$ hypersurfaces in advanced Eddington--Finkelstein-type coordinates (right diagram). While the constant-$t$ hypersurfaces become time-like inside the black hole in region II, the constant-$\tau$ hypersurfaces are always space-like and smoothly continued across the horizons. We now express the Carter tetrad (\ref{CT}) and its dual (\ref{DCT}) in terms of the advanced Eddington--Finkelstein-type coordinates (\ref{AEFTCT}) and (\ref{AEFTCA}), yielding
\begin{equation*}
\begin{split}
\boldsymbol{l}' & = \frac{1}{\sqrt{2 \, \Sigma \, |\Delta|}} \bigl([\Delta + 4 M r] \, \partial_{\tau} + \Delta \, \partial_r + 2 a \, \partial_{\phi}\bigr) \\ \\
\boldsymbol{n}' & = \sqrt{\frac{|\Delta|}{2 \, \Sigma}} (\partial_{\tau} - \partial_r) \\ \\
\boldsymbol{m}' & = \frac{1}{\sqrt{2 \, \Sigma}} \bigl(\textnormal{i} a \sin{(\theta)} \, \partial_{\tau} + \partial_{\theta} + \textnormal{i} \csc{(\theta)} \, \partial_{\phi}\bigr) \\ \\
\boldsymbol{\overline{m}}\hspace{0.03cm}' & = - \frac{1}{\sqrt{2 \, \Sigma}} \bigl(\textnormal{i} a \sin{(\theta)} \, \partial_{\tau} - \partial_{\theta} + \textnormal{i} \csc{(\theta)} \, \partial_{\phi}\bigr)
\end{split}
\end{equation*}
and
\begin{equation*}
\begin{split}
\boldsymbol{l}'_{\textnormal{D}} & = \sqrt{\frac{|\Delta|}{2 \, \Sigma}} \,\, \textnormal{sign}(\Delta) \biggl(\textnormal{d}\tau + \biggl[1 - \frac{2 \Sigma}{\Delta}\biggr] \textnormal{d}r - a \sin^2{(\theta)} \, \textnormal{d}\phi\biggr) \\ \\
\boldsymbol{n}'_{\textnormal{D}} & = \sqrt{\frac{|\Delta|}{2 \, \Sigma}} \bigl(\textnormal{d}\tau + \textnormal{d}r - a \sin^2{(\theta)} \, \textnormal{d}\phi\bigr) \\ \\
\boldsymbol{m}'_{\textnormal{D}} & = \frac{1}{\sqrt{2 \, \Sigma}} \bigl(\textnormal{i} a \sin{(\theta)} \, [\textnormal{d}\tau + \textnormal{d}r] - \Sigma \, \textnormal{d}\theta - \textnormal{i} \, [r^2 + a^2] \sin{(\theta)} \, \textnormal{d}\phi\bigr) \\ \\
\boldsymbol{\overline{m}}\hspace{0.03cm}'_{\textnormal{D}} & = - \frac{1}{\sqrt{2 \, \Sigma}} \bigl(\textnormal{i} a \sin{(\theta)} \, [\textnormal{d}\tau + \textnormal{d}r] + \Sigma \, \textnormal{d}\theta - \textnormal{i} \, [r^2 + a^2] \sin{(\theta)} \,  \textnormal{d}\phi\bigr) \, .
\end{split}
\end{equation*}
Since the real-valued Newman--Penrose vector $\boldsymbol{l}'$ and its dual $\boldsymbol{l}'_{\textnormal{D}}$ are both still singular at the horizons, we apply another class III local Lorentz transformation (\ref{CLIIILLTT}) with parameters
\begin{equation*}
\varsigma = \frac{\sqrt{|\Delta|}}{r_+} \,\,\,\,\, \textnormal{and} \,\,\,\,\, \psi = 0 \, ,
\end{equation*}
which leads to the regular Carter tetrad  
\begin{equation} \label{RCT}
\begin{split}
\boldsymbol{l}'' & = \frac{1}{\sqrt{2 \, \Sigma} \, r_+} \bigl([\Delta + 4 M r] \, \partial_{\tau} + \Delta \, \partial_r + 2 a \, \partial_{\phi}\bigr) \\ \\
\boldsymbol{n}'' & = \frac{r_+}{\sqrt{2 \, \Sigma}} (\partial_{\tau} - \partial_r) \\ \\
\boldsymbol{m}'' & = \frac{1}{\sqrt{2 \, \Sigma}} \bigl(\textnormal{i} a \sin{(\theta)} \, \partial_{\tau} + \partial_{\theta} + \textnormal{i} \csc{(\theta)} \, \partial_{\phi}\bigr) \\ \\
\boldsymbol{\overline{m}}\hspace{0.03cm}'' & = - \frac{1}{\sqrt{2 \, \Sigma}} \bigl(\textnormal{i} a \sin{(\theta)} \, \partial_{\tau} - \partial_{\theta} + \textnormal{i} \csc{(\theta)} \, \partial_{\phi}\bigr)
\end{split}
\end{equation}
and to the dual co-tetrad
\begin{equation*} \label{RDCT}
\begin{split}
\boldsymbol{l}''_{\textnormal{D}} & = \frac{\Delta}{\sqrt{2 \, \Sigma} \, r_+} \biggl(\textnormal{d}\tau + \biggl[1 - \frac{2 \Sigma}{\Delta}\biggr] \textnormal{d}r - a \sin^2{(\theta)} \, \textnormal{d}\phi\biggr) \\ \\
\boldsymbol{n}''_{\textnormal{D}} & = \frac{r_+}{\sqrt{2 \, \Sigma}} \bigl(\textnormal{d}\tau + \textnormal{d}r - a \sin^2{(\theta)} \, \textnormal{d}\phi\bigr) 
\end{split}
\end{equation*}
\begin{equation*}
\begin{split}
\boldsymbol{m}''_{\textnormal{D}} & = \frac{1}{\sqrt{2 \, \Sigma}} \bigl(\textnormal{i} a \sin{(\theta)} \, [\textnormal{d}\tau + \textnormal{d}r] - \Sigma \, \textnormal{d}\theta - \textnormal{i} \, [r^2 + a^2] \sin{(\theta)} \, \textnormal{d}\phi\bigr) \\ \\
\boldsymbol{\overline{m}}\hspace{0.03cm}''_{\textnormal{D}} & = - \frac{1}{\sqrt{2 \, \Sigma}} \bigl(\textnormal{i} a \sin{(\theta)} \, [\textnormal{d}\tau + \textnormal{d}r] + \Sigma \, \textnormal{d}\theta - \textnormal{i} \, [r^2 + a^2] \sin{(\theta)} \, \textnormal{d}\phi\bigr) \, .
\end{split}
\end{equation*}
Substituting the latter into -- and solving -- the torsion-free Maurer--Cartan equation of structure (\ref{NPMCE}), we obtain regular spin coefficients for the non-extreme Kerr geometry in horizon-penetrating advanced Eddington--Finkelstein-type coordinates 
\begin{equation}\label{RENSC}
\begin{split}
\kappa'' & = \sigma'' = \lambda'' = \nu'' = 0 \, , \quad \pi'' = - \tau'' = \frac{\textnormal{i} a \sin{(\theta)}}{\sqrt{2 \, \Sigma} \, \bigl(r - \textnormal{i} a \cos{(\theta)}\bigr)} \, , \quad \mu'' = - \frac{r_+}{\sqrt{2 \, \Sigma} \, \bigl(r - \textnormal{i} a \cos{(\theta)}\bigr)} \, , \\ \\
\varrho'' & = - \frac{\Delta}{\sqrt{2 \, \Sigma} \, r_+ \bigl(r - \textnormal{i} a \cos{(\theta)}\bigr)} \, , \quad \gamma'' = - \frac{r_+}{2^{3/2} \, \sqrt{\Sigma} \, \bigl(r - \textnormal{i} a \cos{(\theta)}\bigr)} \, , \quad \epsilon'' = \frac{r^2 - a^2 - 2 \textnormal{i} a \cos{(\theta)} \, (r - M)}{2^{3/2} \, \sqrt{\Sigma} \, r_+ \bigl(r - \textnormal{i} a \cos{(\theta)}\bigr)} \, , \\ \\
\alpha'' & = - \beta'' = - \frac{1}{(2 \, \Sigma)^{3/2}} \bigl([r^2 + a^2] \cot{(\theta)} - \textnormal{i} r a \sin{(\theta)}\bigr) \, .
\end{split}
\end{equation}

%-----------------------------------------------------------------------
\section{The Massive Dirac Equation in the Analytically Extended Kerr Geometry}
%-----------------------------------------------------------------------

\noindent We determine the explicit form of the massive Dirac equation in the non-extreme Kerr geometry in horizon-penetrating advanced Eddington--Finkelstein-type coordinates using the Newman--Penrose formalism. We show its separability into first-order radial and angular ODE systems, perform asymptotic analyses of the radial solution at infinity, the event horizon, and the Cauchy horizon, and discuss the set of angular eigenfunctions as well as the corresponding eigenvalue spectrum.

%---------------------------------------------------------------------------------------
\subsection{Chandrasekhar's Mode Analysis and Separation of Variables} \label{Sec3}
%---------------------------------------------------------------------------------------

\noindent Inserting the regular Carter tetrad (\ref{RCT}) and the associated spin coefficients (\ref{RENSC}) into the massive Dirac equation (\ref{NPCSTDE}) and employing the mode ansatz (see, e.g., \cite{ChandraBook})
\begin{equation} \label{chandramodeansatz}
\begin{split}
\mathscr{F}_i(\tau, r, \theta, \phi) & =  e^{- \textnormal{i} (\omega \tau + k \phi)} \, \bigl(r - \textnormal{i} a \cos{(\theta)}\bigr)^{-1/2} \, \mathscr{H}_i(r, \theta) \\ 
\mathscr{G}_i(\tau, r, \theta, \phi) & = e^{- \textnormal{i} (\omega \tau + k \phi)} \, \bigl(r + \textnormal{i} a \cos{(\theta)}\bigr)^{-1/2} \, \mathscr{J}_i(r, \theta) \, ,
\end{split}
\end{equation}
where $\omega \in \mathbb{R}$ is the frequency, $k \in \mathbb{Z} + 1/2$ the wave number, and $i \in \{1, 2\}$, we obtain the system of first-order partial differential equations
\begin{equation*}
\begin{split}
& r_+^{- 1} \bigl(\Delta \, \partial_r + r - M - \textnormal{i} \omega \, (\Delta + 4 M r) - 2 \textnormal{i} a k\bigr) \mathscr{H}_1 + \bigl(\partial_{\theta} + 2^{- 1} \cot{(\theta)} - a \omega \sin{(\theta)} - k \csc{(\theta)}\bigr) \mathscr{H}_2 \\  
& =  \textnormal{i} m \bigl(r - \textnormal{i} a \cos{(\theta)}\bigr) \mathscr{J}_1 \\ \\
& r_+ (\partial_r + \textnormal{i} \omega) \mathscr{H}_2 - \bigl(\partial_{\theta} + 2^{- 1} \cot{(\theta)} + a \omega \sin{(\theta)} + k \csc{(\theta)}\bigr) \mathscr{H}_1 = -  \textnormal{i} m \bigl(r - \textnormal{i} a \cos{(\theta)}\bigr) \mathscr{J}_2 \\ \\
& r_+^{- 1} \bigl(\Delta \, \partial_r + r - M - \textnormal{i} \omega \, (\Delta + 4 M r) - 2 \textnormal{i} a k\bigr) \mathscr{J}_2 - \bigl(\partial_{\theta} + 2^{- 1} \cot{(\theta)} + a \omega \sin{(\theta)} + k \csc{(\theta)}\bigr) \mathscr{J}_1 \\
& = \textnormal{i} m \bigl(r + \textnormal{i} a \cos{(\theta)}\bigr) \mathscr{H}_2 \\ \\
& r_+ (\partial_r + \textnormal{i} \omega) \mathscr{J}_1 + \bigl(\partial_{\theta} + 2^{- 1} \cot{(\theta)} - a \omega \sin{(\theta)} - k \csc{(\theta)}\bigr) \mathscr{J}_2 = - \textnormal{i} m \bigl(r + \textnormal{i} a \cos{(\theta)}\bigr) \mathscr{H}_1  
\end{split}
\end{equation*}
for the functions $\mathscr{H}_i$ and $\mathscr{J}_i$, which depends only on the radial and polar angle variables. This system is separable by means of the product ansatz 

\begin{equation*}
\begin{split}
\mathscr{H}_1(r, \theta) & = \mathscr{R}_+(r) \, \mathscr{T}_+(\theta) \\ 
\mathscr{H}_2(r, \theta) & = \mathscr{R}_-(r) \, \mathscr{T}_-(\theta) \\ 
\mathscr{J}_1(r, \theta) & = \mathscr{R}_-(r) \, \mathscr{T}_+(\theta) \\ 
\mathscr{J}_2(r, \theta) & = \mathscr{R}_+(r) \, \mathscr{T}_-(\theta) \, ,
\end{split}
\end{equation*}
yielding the first-order radial ODE system 
\begin{equation*} 
\begin{split}
& \bigl(\Delta \, \partial_r + r - M - \textnormal{i} \omega \, (\Delta + 4 M r) - 2 \textnormal{i} a k\bigr) \, \mathscr{R}_+ = r_+ (\xi + \textnormal{i} m r) \, \mathscr{R}_- \\ 
& r_+ (\partial_r + \textnormal{i} \omega) \, \mathscr{R}_- = (\xi - \textnormal{i} m r) \, \mathscr{R}_+ 
\end{split}
\end{equation*}
and the first-order angular ODE system
\begin{equation*} 
\begin{split}
\bigl(\partial_{\theta} + 2^{- 1} \cot{(\theta)} - a \omega \sin{(\theta)} - k \csc{(\theta)}\bigr) \mathscr{T}_- & = - \bigl(\xi - m a \cos{(\theta)}\bigr) \mathscr{T}_+ \\ 
\bigl(\partial_{\theta} + 2^{- 1} \cot{(\theta)} + a \omega \sin{(\theta)} + k \csc{(\theta)}\bigr) \mathscr{T}_+ & = \bigl(\xi + m a \cos{(\theta)}\bigr) \mathscr{T}_- 
\end{split}
\end{equation*}
with $\xi$ being the constant of separation. Using the functions $\widetilde{\mathscr{R}}_+ := \sqrt{|\Delta|} \, \mathscr{R}_+$ and $\widetilde{\mathscr{R}}_- := r_+ \, \mathscr{R}_-$, the radial system may be transformed into the more symmetric form 
\begin{equation} \label{RADEQ21}
\begin{split}
& \bigl(\Delta \, \partial_r - \textnormal{i} \omega \, (\Delta + 4 M r) - 2 \textnormal{i} a k\bigr) \, \widetilde{\mathscr{R}}_+ = \sqrt{|\Delta|} \, (\xi + \textnormal{i} m r) \, \widetilde{\mathscr{R}}_-  \\ 
& \Delta (\partial_r + \textnormal{i} \omega) \, \widetilde{\mathscr{R}}_- = \textnormal{sign}(\Delta) \sqrt{|\Delta|} \, (\xi - \textnormal{i} m r) \, \widetilde{\mathscr{R}}_+ \, .
\end{split}
\end{equation}
For the analysis of the radial asymptotics at infinity, the event horizon, and the Cauchy horizon, it is, however, advantageous to work with the matrix representation
\begin{equation} \label{Tradeq0}
\partial_r \widetilde{\mathscr{R}} = U(r) \, \widetilde{\mathscr{R}} \, ,
\end{equation}
where $\widetilde{\mathscr{R}} := \bigl(\widetilde{\mathscr{R}}_+, \widetilde{\mathscr{R}}_-\bigr)^{\textnormal{T}}$ and 
\begin{equation*}
U(r) := \frac{1}{\Delta} \left(\begin{array}{cc}
\textnormal{i} \, \bigl(\omega (\Delta + 4 M r) + 2 a k\bigr) & \sqrt{|\Delta|} \, (\xi + \textnormal{i} m r) \\ \\
\textnormal{sign}(\Delta) \sqrt{|\Delta|} \, (\xi - \textnormal{i} m r) & - \textnormal{i} \omega \, \Delta \\
\end{array}\right) .
\end{equation*}
Furthermore, in this representation, the singular points of the radial system can be derived straightforwardly \cite{CoddLevi}. We find that $U$ has singularities of rank $\mu = 0$ at $r = r_{\pm}$, that is, the event and the Cauchy horizon are regular singular points of (\ref{RADEQ21}), even though, there, both the coordinate system and the tetrad frame are non-singular. As a consequence, the regions outside the event horizon $r_+ < r$, between the event and the Cauchy horizon $r_- < r < r_+$, and inside the Cauchy horizon $r < r_-$ have to be considered separately. We emphasize that this is a particularity of the specific mode ansatz (\ref{chandramodeansatz}). Besides, the matrix representation of the angular system is given by 
\begin{equation}\label{MRAE}
A(\theta) \mathscr{T} = \xi \mathscr{T} 
\end{equation}
with $\mathscr{T} := \left(\mathscr{T}_+, \mathscr{T}_-\right)^{\textnormal{T}}$ and
\begin{equation*}
A(\theta) := \left(\begin{array}{cc}
m a \cos{(\theta)} & - \partial_{\theta} - 2^{- 1} \cot{(\theta)} + a \omega \sin{(\theta)} + k \textnormal{csc}(\theta) \\ \\
\partial_{\theta} + 2^{- 1} \cot{(\theta)} + a \omega \sin{(\theta)} + k \textnormal{csc}(\theta) & - m a \cos{(\theta)} 
\end{array}\right) .
\end{equation*}

%-------------------------------------------------------------------------------
\subsection{Asymptotic Analysis of the Radial Solution at Infinity} \label{sec3b}
%-------------------------------------------------------------------------------

\noindent Following the approach of \cite{FKSM3}, we derive the asymptotic solution of the radial system (\ref{Tradeq0}) for $r \rightarrow \infty$ and examine the decay of the associated error. This solution, in addition to the radial asymptotics at the event and the Cauchy horizon, can be used to describe the scattering process of Dirac particles by the gravitational field of a rotating Kerr black hole (cf.\ Section \ref{Sec4}) and for the construction of an integral spectral representation of the Dirac propagator \cite{RF}, both in horizon-penetrating coordinates. We begin by expressing the radial system (\ref{Tradeq0}) in terms of the Regge--Wheeler coordinate 
\begin{equation} \label{Tradeq}
\partial_{r_{\star}} \widetilde{\mathscr{R}} = T(r_{\star}) \, \widetilde{\mathscr{R}} \, ,
\end{equation}
where $T(r_{\star}) := \Delta/(r^2 + a^2) \, U(r_{\star})$. Employing the diagonal matrix $S := D^{- 1} T D = \textnormal{diag}(\lambda_1, \lambda_2)$, with $D$ being the diagonalization matrix and $\lambda_{1/2}$ the eigenvalues of $T$, this system can be written in the form 
\begin{equation*}
\partial_{r_{\star}} \bigl(D^{- 1} \widetilde{\mathscr{R}} \, \bigr) = \bigl[S - D^{- 1}(\partial_{r_{\star}} D)\bigr] \bigl(D^{- 1} \widetilde{\mathscr{R}} \, \bigr) \, .  
\end{equation*}
Hence, using the ansatz 
\begin{equation*}\label{ANS}
\widetilde{\mathscr{R}}(r_{\star}) = D(r_{\star}) \left(\begin{array}{c}
e^{\textnormal{i} \Phi_+(r_{\star})} \, \mathfrak{f}_1(r_{\star}) \\ 
e^{- \textnormal{i} \Phi_-(r_{\star})} \, \mathfrak{f}_2(r_{\star})
\end{array}\right) ,
\end{equation*}
we obtain an ODE system for $\boldsymbol{\mathfrak{f}} := (\mathfrak{f}_1, \mathfrak{f}_2)^{\textnormal{T}}$ given by
\begin{equation*}
\partial_{r_{\star}} \boldsymbol{\mathfrak{f}} = \bigl[S - W^{- 1} D^{- 1} (\partial_{r_{\star}} D) \, W - W^{- 1} \partial_{r_{\star}} W \, \bigr] \boldsymbol{\mathfrak{f}} \, ,
\end{equation*}
where $W := \textnormal{diag}(e^{\textnormal{i} \Phi_+}, e^{- \textnormal{i} \Phi_-})$. We determine the functions $\Phi_{\pm}$ by imposing the condition $S =  W^{- 1} \partial_{r_{\star}} W$, i.e., 
\begin{equation}\label{defrelev}
\partial_{r_{\star}} \Phi_+ = - \textnormal{i} \lambda_1 \quad \textnormal{and} \quad \partial_{r_{\star}} \Phi_- = \textnormal{i} \lambda_2 \, ,
\end{equation}
which yields
\begin{equation}\label{fODE}
\partial_{r_{\star}} \boldsymbol{\mathfrak{f}} = - W^{- 1} D^{- 1} (\partial_{r_{\star}} D) \, W \, \boldsymbol{\mathfrak{f}} \, .
\end{equation}
\begin{Lemma} 
Every nontrivial solution $\widetilde{\mathscr{R}}$ of (\ref{Tradeq}) for $|\omega| > m$ is asymptotically as $r \rightarrow \infty$ of the oscillatory form
\begin{equation} \label{radsolinf}
\widetilde{\mathscr{R}}(r_{\star}) = \widetilde{\mathscr{R}}_{\infty}(r_{\star}) + E_{\infty}(r_{\star}) = D_{\infty} \left(\begin{array}{c}
\mathfrak{f}_{\infty}^{(1)} \, e^{\textnormal{i} \phi_+(r_{\star})} \\ 
\mathfrak{f}_{\infty}^{(2)} \, e^{- \textnormal{i} \phi_-(r_{\star})}
\end{array}\right) + E_{\infty}(r_{\star}) \, ,
\end{equation}	
where 	
\begin{equation} \label{asympdiag}
D_{\infty} := 
\left(\begin{array}{cc}
\cosh{(\Omega)} & \sinh{(\Omega)} \\ 
\sinh{(\Omega)} & \cosh{(\Omega)}
\end{array}\right) 
\quad \textnormal{with} \quad
\Omega := \frac{1}{4} \ln{\biggl(\frac{\omega - m}{\omega + m}\biggr)} \, ,  
\end{equation}
the functions 
\begin{equation} \label{asypha}
\phi_{\pm}(r_{\star}) := \textnormal{sign}(\omega) \biggl[- \sqrt{\omega^2 - m^2} \, r_{\star} + M \biggl(\pm \, 2 \omega - \frac{m^2}{\sqrt{\omega^2 - m^2}}\biggr) \ln{(r_{\star})}\biggr] 
\end{equation}
are the asymptotic phases, and $\boldsymbol{\mathfrak{f}}_{\infty} := \bigl(\mathfrak{f}_{\infty}^{(1)}, \mathfrak{f}_{\infty}^{(2)}\bigr)^{\textnormal{T}} \not= \boldsymbol{0}$ is a vector-valued constant. The error $E_{\infty}$ has polynomial decay
\begin{equation} \label{Rerr}
\|E_{\infty}(r_{\star})\| \leq \frac{a}{r_{\star}} 
\end{equation}
for a suitable constant $a \in \mathbb{R}_{> 0}$. In the case $|\omega| < m$, the non-trivial solution $\widetilde{\mathscr{R}}$ has both contributions that show exponential decay $\sim e^{- \sqrt{m^2 - \omega^2} \, r_{\star}}$ and exponential growth $\sim e^{\sqrt{m^2 - \omega^2} \, r_{\star}}$.
\end{Lemma}
\Proof 
In the limit $r \rightarrow \infty$, the matrix $T$ defined in (\ref{Tradeq}) converges to 
\begin{equation} \label{Tinfty}
T_{\infty} := \lim_{r_{\star} \rightarrow \infty} T = \textnormal{i} \left(\begin{array}{cc}
\omega & m \\ 
- m & - \omega
\end{array}\right).
\end{equation}
Moreover, it has a regular expansion in powers of $1/r_{\star}$, i.e., $T = T_{\infty} + \mathcal{O}(1/r_{\star})$. Accordingly, both the diagonal matrix $S$ and the diagonalization matrix $D$ also have regular expansions in powers of $1/r_{\star}$. 
The eigenvalues of (\ref{Tinfty}) read 
\begin{equation} \label{zoev}
\lambda_{1/2}^{(0)} = \mp \, \textnormal{sign}(\omega) \times
\begin{cases}
\textnormal{i} \sqrt{\omega^2 - m^2} \in \mathbb{C} & \,\, \textnormal{for} \,\,\,\, |\omega| > m \\ 
\hspace{0.1cm} \sqrt{m^2 - \omega^2}  \in \mathbb{R} & \,\, \textnormal{for} \,\,\,\, 
|\omega| < m \, . 
\end{cases}
\end{equation}
For $|\omega| > m$, the diagonalization matrix $D_{\infty} := \lim_{r_{\star} \rightarrow \infty} D$  associated with (\ref{Tinfty}) is given by expression (\ref{asympdiag}). This can be easily verified by direct calculation. Furthermore, with the asymptotic first-order eigenvalues of $T$ 
\begin{equation*}
\lambda_{1/2}^{(1)} = \textnormal{sign}(\omega) \biggl[\mp \, \textnormal{i} \sqrt{\omega^2 - m^2} + \displaystyle{\frac{\textnormal{i} M}{r_{\star}} \left(2 \omega \mp \frac{m^2}{\sqrt{\omega^2 - m^2}}\right)}\biggr] \, ,
\end{equation*}
we can solve (\ref{defrelev}) by simple integration and obtain (\ref{asypha}) as the asymptotic phases. We point out that for the determination of the asymptotic phases, it is of paramount importance to take the first-order terms of the asymptotic eigenvalues into account because they yield contributions to the radial solution that do not decay at infinity. However, the first-order terms play no role in the computation of $D_{\infty}$, as the corresponding contributions can be absorbed into the error $E_{\infty}$. Next, since for $r_{\star}$ sufficiently close to infinity the Hilbert--Schmidt norms of $D^{- 1}$ and $\partial_{r_{\star}} D$ are bounded from above by
\begin{equation*}
\|D^{- 1}\|_{\textnormal{HS}} \leq c \quad \textnormal{and} \quad \|\partial_{r_{\star}} D\|_{\textnormal{HS}} \leq \frac{d}{r^2_{\star}} \, ,
\end{equation*}
where both $c$ and $d$ denote positive constants, the $\mathbb{C}^2$-norm of (\ref{fODE}) may be estimated by
\begin{equation} \label{fest}
\|\partial_{r_{\star}} \boldsymbol{\mathfrak{f}}\| \leq 2 \, \|D^{- 1}\|_{\textnormal{HS}} \cdot \|\partial_{r_{\star}} D\|_{\textnormal{HS}} \cdot \|\boldsymbol{\mathfrak{f}}\| \leq \frac{2 c d}{r^2_{\star}} \, \|\boldsymbol{\mathfrak{f}}\|
\end{equation}
with $\|W\|_{\textnormal{HS}} = \|W^{- 1}\|_{\textnormal{HS}} = \sqrt{2}$. Applying the triangle and the Cauchy--Schwarz inequality, one can derive the following inequality 
\begin{equation*}
\begin{split}
\bigl|\partial_{r_{\star}} \|\boldsymbol{\mathfrak{f}}\|\bigr| = \frac{\left|\partial_{r_{\star}} \left\langle\boldsymbol{\mathfrak{f}}, \boldsymbol{\mathfrak{f}}\right\rangle\right|}{2 \, \|\boldsymbol{\mathfrak{f}}\|} = \frac{\left|\left\langle\boldsymbol{\mathfrak{f}}, \partial_{r_{\star}} \boldsymbol{\mathfrak{f}}\right\rangle + \left\langle\partial_{r_{\star}} \boldsymbol{\mathfrak{f}}, \boldsymbol{\mathfrak{f}}\right\rangle\right|}{2 \, \|\boldsymbol{\mathfrak{f}}\|} \leq \frac{\left|\left\langle\boldsymbol{\mathfrak{f}}, \partial_{r_{\star}} \boldsymbol{\mathfrak{f}}\right\rangle\right| + \left|\left\langle\partial_{r_{\star}} \boldsymbol{\mathfrak{f}}, \boldsymbol{\mathfrak{f}}\right\rangle\right|}{2 \, \|\boldsymbol{\mathfrak{f}}\|} = \frac{\left|\left\langle\boldsymbol{\mathfrak{f}}, \partial_{r_{\star}} \boldsymbol{\mathfrak{f}}\right\rangle\right|}{\|\boldsymbol{\mathfrak{f}}\|} \leq \frac{\|\boldsymbol{\mathfrak{f}}\| \cdot \|\partial_{r_{\star}} \boldsymbol{\mathfrak{f}}\|}{\|\boldsymbol{\mathfrak{f}}\|} = \|\partial_{r_{\star}} \boldsymbol{\mathfrak{f}}\| \, .
\end{split}
\end{equation*}
Using this inequality in (\ref{fest}), we find
\begin{equation}\label{fHS}
\bigr|\partial_{r_{\star}} \|\boldsymbol{\mathfrak{f}}\| \bigr| \leq \frac{2 c d}{r^2_{\star}} \, \|\boldsymbol{\mathfrak{f}}\| \, .
\end{equation}
We note that $\|\boldsymbol{\mathfrak{f}}\| \not= 0$ because $\widetilde{\mathscr{R}}$ has to be nontrivial. Now, integrating (\ref{fHS}) with respect to the Regge--Wheeler coordinate from $r_0$ to $r_{\star}$ and employing the triangle inequality for integrals gives for all $0 < r_0 \leq r_{\star}$
\begin{equation*}
\left|\int_{r_0}^{r_{\star}} \partial_{r'_{\star}} \ln{\|\boldsymbol{\mathfrak{f}}\|} \, \textnormal{d}r'_{\star}\right| \leq \int_{r_0}^{r_{\star}} \left|\partial_{r'_{\star}} \ln{\|\boldsymbol{\mathfrak{f}}\|}\right| \textnormal{d}r'_{\star} \leq 2 c d \int_{r_0}^{r_{\star}} \frac{\textnormal{d}r'_{\star}}{r'^2_{\star}} \, ,
\end{equation*}
and hence
\begin{equation*}
\left|\ln{\|\boldsymbol{\boldsymbol{\mathfrak{f}}}\|}\Bigl|^{r_{\star}}_{r_0} \right| \leq - \frac{2 c d}{r'_{\star}} \biggl|^{r_{\star}}_{r_0} \, .
\end{equation*}
Consequently, since $0 < 2 c d/r'_{\star}\bigl|^{r_0}_{r_{\star}} < \infty$, there exists a constant $N > 0$ such that  
\begin{equation} \label{NNBOUND}
\frac{1}{N} \leq \|\boldsymbol{\mathfrak{f}}\| \leq N \, .
\end{equation}
Combining this with (\ref{fest}), we obtain for sufficiently large values of $r_{\star}$
\begin{equation}\label{fest2}
\|\partial_{r_{\star}} \boldsymbol{\mathfrak{f}}\| \leq \frac{b}{r^2_{\star}} \, ,
\end{equation}
where $b := 2 c d N$, which implies that $\boldsymbol{\mathfrak{f}}$ is integrable and, according to (\ref{NNBOUND}), has a finite, non-zero limit $\boldsymbol{\mathfrak{f}}_{\infty}:= \lim_{r_{\star} \rightarrow \infty} \boldsymbol{\mathfrak{f}}(r_{\star}) \not= \boldsymbol{0}$. Integrating (\ref{fest2}) from $r_{\star}$ to $\infty$ and again using the triangle inequality for integrals yields the error estimate
\begin{equation}\label{fest3}
\|E_{\boldsymbol{\mathfrak{f}}}\| = \|\boldsymbol{\mathfrak{f}} - \boldsymbol{\boldsymbol{\mathfrak{f}}}_{\infty}\| = \left\|\int^{\infty}_{r_{\star}} \partial_{r'_{\star}} \boldsymbol{\mathfrak{f}} \, \textnormal{d}r'_{\star}\right\| \leq \int^{\infty}_{r_{\star}} \|\partial_{r'_{\star}} \boldsymbol{\mathfrak{f}}\| \, \textnormal{d}r'_{\star} \leq \frac{b}{r_{\star}} \, .
\end{equation}
The polynomial $1/r_{\star}$-decay of the error $E_{\infty}$ in (\ref{Rerr}) follows directly from the estimate (\ref{fest3}) and the fact that the matrices $T$ and $D$ -- and therefore their eigenvalues -- have regular expansions in powers of $1/r_{\star}$. This can be verified via the following short computation. Inserting the expansion of $D$ and of the eigenvalues of $T$ into the $\mathbb{C}^2$-norm of the error $E_{\infty}$, we find
\begin{equation*}
\begin{split}
\|E_{\infty}\| & = \bigl\|\widetilde{\mathscr{R}} - \widetilde{\mathscr{R}}_{\infty}\bigr\| = \left\|D \left(\begin{array}{c}
e^{\textnormal{i} \Phi_+} \, \mathfrak{f}_1 \\ 
e^{- \textnormal{i} \Phi_-} \, \mathfrak{f}_2
\end{array}\right) - D_{\infty} \left(\begin{array}{l}
\mathfrak{f}_{\infty}^{(1)} \, e^{\textnormal{i} \phi_+} \\ 
\mathfrak{f}_{\infty}^{(2)} \, e^{- \textnormal{i} \phi_-}
\end{array}\right)\right\| \\ \\
& = \left\|\biggl[D_{\infty} + \mathcal{O}\biggl(\frac{1}{r_{\star}}\biggr)\biggr] \left(\begin{array}{c}
e^{\textnormal{i} \phi_+ + \mathcal{O}(1/r_{\star})} \, \mathfrak{f}_1 \\ 
e^{- \textnormal{i} \phi_- + \mathcal{O}(1/r_{\star})} \, \mathfrak{f}_2
\end{array}\right) - D_{\infty} \left(\begin{array}{l}
\mathfrak{f}_{\infty}^{(1)} \, e^{\textnormal{i} \phi_+} \\ 
\mathfrak{f}_{\infty}^{(2)} \, e^{- \textnormal{i} \phi_-} 
\end{array}\right)\right\| .
\end{split}
\end{equation*}
Subsequently, we collect terms of the order $\mathcal{O}(1/r_{\star})$ and apply the triangle inequality as well as the submultiplicativity property of operator norms, which results in the estimate
\begin{equation*}
\|E_{\infty}\| = \left\|D_{\infty} W_{\infty} \, (\boldsymbol{\mathfrak{f}} - \boldsymbol{\mathfrak{f}}_{\infty}) + \mathcal{O}\biggl(\frac{1}{r_{\star}}\biggr)\right\| \leq \left\|D_{\infty}\right\|_{\textnormal{HS}} \cdot \left\|W_{\infty}\right\|_{\textnormal{HS}} \cdot \|E_{\boldsymbol{\mathfrak{f}}}\| + \left\|\mathcal{O}\biggl(\frac{1}{r_{\star}}\biggr)\right\| ,
\end{equation*}
where $W_{\infty} := \textnormal{diag}(e^{\textnormal{i} \phi_+}, e^{- \textnormal{i} \phi_-})^{\textnormal{T}}$. Since $D_{\infty}$ is a constant, non-zero matrix, the associated Hilbert--Schmidt norm yields a positive constant. Moreover, one immediately verifies that $\left\|W_{\infty}\right\|_{\textnormal{HS}} = \sqrt{2}$. From this and (\ref{fest3}), we can thus conclude that the error $E_{\infty}$ has the polynomial decay $\|E_{\infty}\| \leq a/r_{\star}$, with $a$ being a positive constant. We note in passing that in the case $|\omega| < m$, it is obvious from the zero-order eigenvalues (\ref{zoev}) that the non-trivial solution $\widetilde{\mathscr{R}}$ has contributions with exponential decay and exponential growth.
\QED

%---------------------------------------------------------------------------------------------
\subsection{Asymptotic Analysis of the Radial Solution at the Event Horizon} \label{sec:4C}
%---------------------------------------------------------------------------------------------

\noindent In order to derive the asymptotics of (\ref{Tradeq}) at the event horizon, we employ the solution ansatz
\begin{equation*}\label{RADANS}
\widetilde{\mathscr{R}} = \left(\begin{array}{c}
e^{2 \textnormal{i} \bigl(\omega + k \Omega^{(+)}_{\textnormal{Kerr}}\bigr) r_{\star}} \, \mathfrak{g}_1(r_{\star}) \\
\mathfrak{g}_2(r_{\star}) \end{array}\right) ,
\end{equation*}
where $\Omega^{(+)}_{\textnormal{Kerr}} := a/(2 M r_+)$ is the angular velocity of the event horizon, obtaining an ODE system for $\boldsymbol{\mathfrak{g}} := (\mathfrak{g}_1, \mathfrak{g}_2)^{\textnormal{T}}$ given by
\begin{equation}\label{geq}
\begin{split}
\partial_{r_{\star}} \boldsymbol{\mathfrak{g}} & = - \frac{\textnormal{i}}{r^2 + a^2} \left[2 k \, \bigl(2 M \Omega^{(+)}_{\textnormal{Kerr}} \, r - a\bigr) 
\left(\begin{array}{cc}
1 & 0 \\ 
0 & 0 \end{array} \right) 
+ \textnormal{sign}(\Delta) \sqrt{|\Delta|} \left.\begin{array}{cc}
&  \\ \\
& \end{array}\right. \right. \\ \\
& \hspace{2.3cm} \left. \times \left(\begin{array}{cc}
\sqrt{|\Delta|} \, \bigl(\omega + 2 k \Omega^{(+)}_{\textnormal{Kerr}}\bigr) & e^{- 2 \textnormal{i} \bigl(\omega + k \Omega^{(+)}_{\textnormal{Kerr}}\bigr) r_{\star}} \, \textnormal{sign}(\Delta) \, (\textnormal{i} \xi - m r) \\ \\
e^{2 \textnormal{i} \bigl(\omega + k \Omega^{(+)}_{\textnormal{Kerr}}\bigr) r_{\star}} (\textnormal{i} \xi + m r) &  \sqrt{|\Delta|} \, \omega \end{array}\right)
\right] \boldsymbol{\mathfrak{g}} \, .
\end{split}
\end{equation}
In the limit $r \searrow r_+$, the right hand side of this system vanishes, and thus we find the asymptotic solution $\boldsymbol{\mathfrak{g}}_{r_+} := \lim_{r \searrow r_+} \boldsymbol{\mathfrak{g}} = \textnormal{const.}$
\begin{Lemma} \label{L4.2}
Every nontrivial solution $\widetilde{\mathscr{R}}$ of (\ref{Tradeq}) is asymptotically as $r \searrow r_+$ of the form
\begin{equation}\label{ntans2}
\widetilde{\mathscr{R}}(r_{\star}) = \widetilde{\mathscr{R}}_{r_+}(r_{\star}) + E_{r_+}(r_{\star}) = \left(\begin{array}{c}
\mathfrak{g}_{r_+}^{(1)} \, e^{2 \textnormal{i} \bigl(\omega + k \Omega^{(+)}_{\textnormal{Kerr}}\bigr) r_{\star}} \\ 
\mathfrak{g}_{r_+}^{(2)}
\end{array}\right) + E_{r_+}(r_{\star}) 
\end{equation}
with
\begin{equation*}
\boldsymbol{\mathfrak{g}}_{r_+} := \bigl(\mathfrak{g}_{r_+}^{(1)}, \mathfrak{g}_{r_+}^{(2)}\bigr)^{\textnormal{T}} = \textnormal{const.} \not= \boldsymbol{0}
\end{equation*}
and an error $E_{r_+}$ with exponential decay
\begin{equation*}
\|E_{r_+}(r_{\star})\| \leq p \, e^{q r_\star}
\end{equation*}
for $r$ sufficiently close to $r_+$ and suitable constants $p, q \in \mathbb{R}_{> 0}$.
\end{Lemma}
\Proof 
For $r \searrow r_+$, as $r \sim r_+ + e^{2 q r_{\star}}$ with $q := (r_+ - r_-)/(4 M r_+) \in \mathbb{R}_{> 0}$, the right hand side of (\ref{geq}) is of the order $\mathcal{O}(e^{q r_{\star}})$. Hence, there exists a constant $p' \in \mathbb{R}_{> 0}$ such that for $r_{\star}$ sufficiently close to $- \infty$
\begin{equation} \label{gbound}
\|\partial_{r_{\star}} \boldsymbol{\mathfrak{g}}\| \leq p' e^{q r_{\star}} \|\boldsymbol{\mathfrak{g}}\| \, .
\end{equation}
From this inequality, following the method of proof of the previous subsection, we can derive the estimate
\begin{equation*}
\left|\ln{\|\boldsymbol{\mathfrak{g}}\|}\Bigl|^{r_0}_{r_{\star}} \right| \leq \frac{p'}{q} \, e^{q r'_{\star}} \Bigl|^{r_0}_{r_{\star}} 
\end{equation*}
for all $r_{\star} \leq r_0$, and since $e^{q r'_{\star}}\bigl|^{r_0}_{r_{\star}}$ is positive, there is a constant $N' > 0$ such that the $\mathbb{C}^2$-norm of $\boldsymbol{\mathfrak{g}}$ is bounded by
\begin{equation*} 
\frac{1}{N'} \leq \|\boldsymbol{\mathfrak{g}}\| \leq N' \, .
\end{equation*}
Combining this bound with inequality (\ref{gbound}) yields
\begin{equation} \label{NSTBOUND}
\|\partial_{r_{\star}} \boldsymbol{\mathfrak{g}}\| \leq p \, e^{q r_{\star}} \, ,
\end{equation}
where $p := p' N'$. Altogether, this implies that $\boldsymbol{\mathfrak{g}}$ is integrable and has a finite, non-zero limit for $r_{\star} \rightarrow - \infty$. Integrating (\ref{NSTBOUND}) with respect to the Regge--Wheeler coordinate from $- \infty$ to $r_{\star}$ and applying the triangle inequality for integrals results in 
\begin{equation}\label{fingb}
\|E_{\boldsymbol{\mathfrak{g}}}\| = \|\boldsymbol{\mathfrak{g}} - \boldsymbol{\mathfrak{g}}_{r_+}\| = \left\|\int_{- \infty}^{r_{\star}} \partial_{r'_{\star}} \boldsymbol{\mathfrak{g}} \, \textnormal{d}r'_{\star}\right\| \leq \int_{- \infty}^{r_{\star}} \|\partial_{r'_{\star}} \boldsymbol{\mathfrak{g}}\| \, \textnormal{d}r'_{\star} \leq p \, e^{q r_{\star}} \, ,
\end{equation} 
which proves the exponential decay of the error $E_{\boldsymbol{\mathfrak{g}}}$. The exponential decay of the error $E_{r_+}$ follows directly from (\ref{fingb}). 
\QED

%------------------------------------------------------------------------------------------------
\subsection{Asymptotic Analysis of the Radial Solution at the Cauchy Horizon} \label{sec:4D}
%------------------------------------------------------------------------------------------------

\noindent Similarly to the analysis of the radial asymptotics at the event horizon, we begin with a solution ansatz of the form
\begin{equation*}\label{RADANS2}
\widetilde{\mathscr{R}} = \left(\begin{array}{c}
e^{2 \textnormal{i} \bigl(\omega + k \Omega^{(-)}_{\textnormal{Kerr}}\bigr) r_{\star}} \, \mathfrak{h}_1(r_{\star}) \\
\mathfrak{h}_2(r_{\star}) \end{array}\right), 
\end{equation*}
where $\Omega^{(-)}_{\textnormal{Kerr}} := a/(2 M r_-)$ is the angular velocity of the Cauchy horizon, and substitute it into the system (\ref{Tradeq}). This leads to an ODE system for $\boldsymbol{\mathfrak{h}} := (\mathfrak{h}_1, \mathfrak{h}_2)^{\textnormal{T}}$ reading
\begin{equation*}\label{heq}
\begin{split}
\partial_{r_{\star}} \boldsymbol{\mathfrak{h}} & = - \frac{\textnormal{i}}{r^2 + a^2} \left[2 k \, \bigl(2 M \Omega^{(-)}_{\textnormal{Kerr}} r - a\bigr) 
\left(\begin{array}{cc}
1 & 0 \\ 
0 & 0 \end{array} \right) 
+ \textnormal{sign}(\Delta) \sqrt{|\Delta|} \left.\begin{array}{cc}
&  \\ \\
& \end{array}\right. \right. \\ \\
& \hspace{2.3cm} \left. \times \left(\begin{array}{cc}
\sqrt{|\Delta|} \, \bigl(\omega + 2 k \Omega^{(-)}_{\textnormal{Kerr}}\bigr) & e^{- 2 \textnormal{i} \bigl(\omega + k \Omega^{(-)}_{\textnormal{Kerr}}\bigr) r_{\star}} \, \textnormal{sign}(\Delta) \, (\textnormal{i} \xi - m r) \\ \\
e^{2 \textnormal{i} \bigl(\omega + k \Omega^{(-)}_{\textnormal{Kerr}}\bigr) r_{\star}} (\textnormal{i} \xi + m r) & \sqrt{|\Delta|} \, \omega \end{array}\right)
\right] \boldsymbol{\mathfrak{h}} \, .
\end{split}
\end{equation*}
In the limit $r \searrow r_-$, the square bracket on the right hand side vanishes, giving rise to the asymptotic radial solution $\boldsymbol{\mathfrak{h}}_{r_-} := \lim_{r \searrow r_-} \boldsymbol{\mathfrak{h}} = \textnormal{const.}$ 
\begin{Lemma} \label{L4.3}
Every nontrivial solution $\widetilde{\mathscr{R}}$ of (\ref{Tradeq}) is asymptotically as $r \searrow r_-$ of the form
\begin{equation*}\label{ntans3}
\widetilde{\mathscr{R}}(r_{\star}) = \widetilde{\mathscr{R}}_{r_-}(r_{\star}) + E_{r_-}(r_{\star}) = \left(\begin{array}{c}
\mathfrak{h}_{r_-}^{(1)} \, e^{2 \textnormal{i} \bigl(\omega + k \Omega^{(-)}_{\textnormal{Kerr}}\bigr) r_{\star}} \\ 
\mathfrak{h}_{r_-}^{(2)}
\end{array}\right) + E_{r_-}(r_{\star}) \, ,
\end{equation*}
with
\begin{equation*}
\boldsymbol{\mathfrak{h}}_{r_-} := \bigl(\mathfrak{h}_{r_-}^{(1)}, \mathfrak{h}_{r_-}^{(2)}\bigr)^{\textnormal{T}} = \textnormal{const.} \not= \boldsymbol{0}
\end{equation*}
and an error $E_{r_-}$ with exponential decay
\begin{equation*}
\|E_{r_-}(r_{\star})\| \leq u \, e^{- v r_\star}
\end{equation*}
for $r$ sufficiently close to $r_-$ and suitable constants $u, v \in \mathbb{R}_{> 0}$.
\end{Lemma}
\Proof The proof of this lemma is identical to the proof of Lemma \ref{L4.2}. \QED

%-----------------------------------------------------------------------------------
\subsection{Angular Eigenfunctions and Eigenvalues} \label{sec:4E}
%-----------------------------------------------------------------------------------

\noindent We discuss the set of eigenfunctions as well as the eigenvalue spectrum of the angular system (\ref{MRAE}), which in its decoupled second-order form is known as the massive Chandrasekhar--Page equation \cite{ChandraBook}
\begin{equation*}
\begin{split}
& \biggl[\partial_{\theta \theta} + \biggl(\cot{(\theta)} \pm \frac{m a \sin{(\theta)}}{\xi \pm m a \cos{(\theta)}}\biggr) \partial_{\theta} - \frac{\csc^2{(\theta)}}{2} \biggl(1 - \frac{3 \cos^2{(\theta)}}{2}\biggr) - 2 k a \omega - \biggl(\frac{\cot{(\theta)}}{2} \mp a \omega \sin{(\theta)}\biggr)^2 \\ \\
& \hspace{0.2cm} - \biggl(\frac{\cot{(\theta)}}{2} \pm k \csc{(\theta)}\biggr)^2 \pm \frac{m a \sin{(\theta)}}{\xi \pm m a \cos{(\theta)}} \biggl(\frac{\cot{(\theta)}}{2} \pm a\omega \sin{(\theta)} \pm k \csc{(\theta)}\biggr) + \xi^2 - m^2 a^2 \cos^2{(\theta)}\biggr] \mathscr{T}_{\pm} = 0 \, .
\end{split}
\end{equation*}
In the limit $a \searrow 0$, the solutions of this equation reduce to the spin-weighted spherical harmonics for the spin-$1/2$ case \cite{GMNRS}. For non-zero angular momenta, the solutions are usually referred to as spin-$1/2$ spheroidal harmonics. A good introduction and a compilation of the basic properties of these functions can be found in the recent paper \cite{DolGair}. For the purposes of the present work and the constructions in \cite{RF}, it is, however, only of importance that the matrix-valued differential operator on the left hand side of (\ref{MRAE}) has a spectral decomposition with smooth eigenfunctions and discrete, non-degenerate eigenvalues, which was already proven in \cite{FKSM1, FKSM3}. In the following, we state the relevant results.
\begin{Prp} \label{PropAEFEV}
For any $\omega \in \mathbb{R}$ and $k \in \mathbb{Z} + 1/2$, the differential operator in \eqref{MRAE} has a complete set of orthonormal eigenfunctions $(\mathscr{T}_n)_{n \in \mathbb{Z}}$ in $L^2 \bigl((0, \pi), \sin{(\theta)} \, \textnormal{d}\theta\bigr)^2$ that are bounded and smooth away from the poles. The corresponding eigenvalues $\xi_n$ are real-valued and non-degenerate, and can thus be ordered as $\xi_n < \xi_{n + 1}$. Both the eigenfunctions and the eigenvalues depend smoothly on $\omega$.
\end{Prp}

%--------------------------------------------------------------------------------------------------------------
\section{Scattering of Massive Dirac Particles by the Gravitational Field of a Rotating Kerr Black Hole}\label{Sec4}
%--------------------------------------------------------------------------------------------------------------

\noindent In this section, we study the scattering of massive Dirac particles by the gravitational field of a rotating Kerr black hole from the point of view of an observer described by horizon-penetrating advanced Eddington--Finkelstein-type coordinates. To this end, we consider Dirac particles that emerge from space-like infinity and propagate toward the black hole's event horizon, where they are either reflected or transmitted. We compute the net current of Dirac particles at infinity as well as at the event horizon, and further derive a condition for the reflection and transmission coefficients. \newline \newline
Using the asymptotic radial solutions at infinity (\ref{radsolinf}) and the event horizon (\ref{ntans2}), we impose boundary conditions specifying an incident wave of unit amplitude at infinity, a reflected wave of amplitude $A_{\omega, m}$ at infinity, and a transmitted wave of amplitude $B_{\omega, m}$ at the event horizon, yielding
\begin{equation}\label{scsolinf}
\widetilde{\mathscr{R}}_{\textnormal{scat.}}(r \rightarrow \infty) \sim D_{\infty} \left(\begin{array}{c}
e^{\textnormal{i} \phi_+(r_{\star})} \\
A_{\omega, m} \, e^{- \textnormal{i} \phi_-(r_{\star})}
\end{array}\right)
\end{equation}
and
\begin{equation}\label{scsoleh}
\widetilde{\mathscr{R}}_{\textnormal{scat.}}(r \searrow r_+) \sim \left(\begin{array}{c}
B_{\omega, m} \, e^{2 \textnormal{i} \bigl(\omega + k \Omega^{(+)}_{\textnormal{Kerr}}\bigr) r_{\star}} \\
0
\end{array}\right).
\end{equation}
We point out that these boundary conditions are chosen in conformity with the physical requirement that classically no particles can emerge from the event horizon. Besides, only the branch $\omega > m$ is considered because free particles at infinity must have energies that exceed -- or at least equal -- their rest energies. Next, assuming the normalization condition 
\begin{equation*}
\|\mathscr{T}_+(\theta)\|^2 + \|\mathscr{T}_-(\theta)\|^2 = 1
\end{equation*}
for the angular eigenfunctions (cf.\ Proposition \ref{PropAEFEV}), the radial Dirac current (for definitions and notations see Appendix \ref{AB}) becomes
\begin{equation*}
J^r = \sqrt{2} \, \sigma\indices{^r_{A B'}} \Bigl(P^A \, \overline{P}^{B'} + Q^A \, \overline{Q}^{B'}\Bigr) = 
\frac{1}{ r_+ \Sigma} \Bigl(\textnormal{sign}{(\Delta)} \, \big\|\widetilde{\mathscr{R}}_{+}\big\|^2 - \big\|\widetilde{\mathscr{R}}_{-}\big\|^2 \Bigr)
\end{equation*}
with the radial Infeld--van der Waerden symbol
\begin{equation*}
\sigma\indices{^r_{A B'}} = \left(\begin{array}{cc}
l^r & m^r \\
\overline{m}^{\, r} & n^r
\end{array}\right)_{A B'} = \frac{1}{\sqrt{2 \, \Sigma} \, r_+}
\, \left(\begin{array}{cc}
\Delta & 0 \\
0 & - r_+^2
\end{array}\right)_{A B'}.
\end{equation*}
As this current has jump discontinuities at the event and the Cauchy horizon (see the final paragraph of Section \ref{Sec3}), it has to be evaluated separately in each of the respective three regions. For the scattering problem at hand, however, it suffices to consider the exterior region of the black hole $r_+ \leq r$, where the radial Dirac current reads
\begin{equation}\label{raddircurr2}
J^r = \frac{1}{ r_+ \Sigma} \Bigl(\big\|\widetilde{\mathscr{R}}_{+}\big\|^2 - \big\|\widetilde{\mathscr{R}}_{-}\big\|^2 \Bigr).
\end{equation}
From the radial system (\ref{Tradeq0}) and its complex conjugate, we obtain the relation
\begin{equation*}
\big\|\widetilde{\mathscr{R}}_{+}\big\|^2 - \big\|\widetilde{\mathscr{R}}_{-}\big\|^2 = \textnormal{const.}
\end{equation*}
via simple algebraic manipulations. Substituting this relation into the radial Dirac current (\ref{raddircurr2}), we derive the conserved net current of Dirac particles
\begin{equation*}
\partial_{\tau} N = \int_0^{2 \pi} \int_0^{\pi} J^r \sqrt{|\textnormal{det}(\boldsymbol{g})|} \,\, \textnormal{d}\theta \, \textnormal{d}\phi = \frac{4 \pi}{r_+} \Bigl(\big\|\widetilde{\mathscr{R}}_{+}\big\|^2 - \big\|\widetilde{\mathscr{R}}_{-}\big\|^2\Bigr) = \textnormal{const.} \, ,
\end{equation*}
where $N$ is the total number of Dirac particles and $\sqrt{|\textnormal{det}(\boldsymbol{g})|} = \Sigma \sin{(\theta)}$. Defining the reflection and transmission coefficients
\begin{equation*}\label{RpRmRel}
R_{\omega, m} := |A_{\omega, m}|^2 \quad \textnormal{and} \quad T_{\omega, m} := |B_{\omega, m}|^2
\end{equation*}
and employing the asymptotic radial solutions (\ref{scsolinf}) and (\ref{scsoleh}), the net current at infinity and at the event horizon results in 
\begin{equation*}
\partial_{\tau} N_{\, |r \rightarrow \infty} = \frac{4 \pi}{r_+} \, (1 - R_{\omega, m})
\end{equation*}
and
\begin{equation*}
\partial_{\tau} N_{\, |r \searrow r_+} = \frac{4 \pi}{r_+} \, T_{\omega, m} \, .
\end{equation*}
From the constancy of the net current and these expressions, we infer that
\begin{equation*}
R_{\omega, m} + T_{\omega, m} = 1 \, ,
\end{equation*}
which proves that superradiance cannot occur because the reflection coefficient is always less than unity. Furthermore, we have shown that the net current across the event horizon is always positive. These results are in agreement with those found in the accordant analysis of this scattering problem performed in terms of Boyer--Lindquist coordinates (see \cite{ChandraBook} and references therein).

\vspace{0.3cm}
 
%------------------------------------------
\section*{Acknowledgments}
%------------------------------------------

\noindent The author is grateful to Felix Finster and Niky Kamran for many useful discussions and comments, as well as for a careful reading this paper. This work was supported by the DFG research grant ``Dirac Waves in the Kerr Geometry: Integral Representations, Mass Oscillation Property and the Hawking Effect.''

\vspace{0.3cm}
  
\begin{appendix}

%------------------------------------------------------------------------
\section{The Newman--Penrose Formalism}\label{AA}
%------------------------------------------------------------------------

\noindent We let $(\mathfrak{M}, \boldsymbol{g})$ be a Lorentzian $4$-manifold endowed with the unique, torsion-free Levi--Civita connection $\boldsymbol{\omega}$ and dual bases $(\boldsymbol{e}_{\mu})$ and $(\boldsymbol{e}^{\mu})$, $\mu \in \{0, 1, 2, 3\}$, on sections of the tangent and cotangent bundles $T\mathfrak{M}$ and $T^{\star}\mathfrak{M}$, respectively. Moreover, we let $F\mathfrak{M}$ and $F^{\star}\mathfrak{M}$ be local (orthonormal or null) frame bundles on $\mathfrak{M}$. The bases on sections of these frame bundles consist of four vectors and four dual co-vectors denoted by $(\boldsymbol{e}_{(a)})$ and $(\boldsymbol{e}^{(a)})$, $a \in \{0, 1, 2, 3\}$. In terms of the former bases of the tangent and cotangent spaces, they can be written as  
\begin{equation*}
\boldsymbol{e}\indices{_{(a)}} = e\indices{^{\mu}_{(a)}} \boldsymbol{e}\indices{_{\mu}} \quad \textnormal{and} \quad \boldsymbol{e}\indices{^{(a)}} = e\indices{_{\mu}^{(a)}} \boldsymbol{e}\indices{^{\mu}} \, , 
\end{equation*}
where $e\indices{^{\mu}_{(a)}}$ is a linear map from $T\mathfrak{M}$ to $F\mathfrak{M}$, namely a ($4 \times 4$)-matrix, and $e\indices{_{\mu}^{(a)}}$ is its inverse. In the Newman--Penrose formalism \cite{NP}, the local bases are composed of two real-valued null vectors, $\boldsymbol{l} = \boldsymbol{e}\indices{_{(0)}} = \boldsymbol{e}\indices{^{(1)}}$ and $\boldsymbol{n} = \boldsymbol{e}\indices{_{(1)}} = \boldsymbol{e}\indices{^{(0)}}$, as well as a complex-conjugate pair of null vectors, $\boldsymbol{m} = \boldsymbol{e}\indices{_{(2)}} = - \boldsymbol{e}\indices{^{(3)}}$ and $\boldsymbol{\overline{m}} = \boldsymbol{e}\indices{_{(3)}} = - \boldsymbol{e}\indices{ ^{(2)}}$ (following the notation and terminology of \cite{ChandraBook}). These are required to satisfy the null conditions
\begin{equation} \label{NPNC}
\boldsymbol{l} \cdot \boldsymbol{l} = \boldsymbol{n} \cdot \boldsymbol{n} = \boldsymbol{m} \cdot \boldsymbol{m} = \boldsymbol{\overline{m}} \cdot \boldsymbol{\overline{m}} = 0 \, ,
\end{equation}
the orthogonality conditions
\begin{equation}\label{NPOC}
\boldsymbol{l} \cdot \boldsymbol{m} = \boldsymbol{l} \cdot \boldsymbol{\overline{m}} = \boldsymbol{n} \cdot \boldsymbol{m} = \boldsymbol{n} \cdot \boldsymbol{\overline{m}} = 0\, ,
\end{equation}
and the cross-normalization conditions (which depend on the signature convention)
\begin{equation} \label{NPCNC}
\boldsymbol{l} \cdot \boldsymbol{n} = - \boldsymbol{m} \cdot \boldsymbol{\overline{m}} = 1\, . 
\end{equation}
The corresponding local metric is non-degenerate and constant, reading
\begin{equation*}
\boldsymbol{\eta} = g\indices{_{\mu \nu}} \, e\indices{^{\mu}_{(a)}} \, e\indices{^{\nu}_{(b)}} \,  \boldsymbol{e}\indices{^{(a)}} \otimes \boldsymbol{e}\indices{^{(b)}} = \boldsymbol{l} \otimes \boldsymbol{n} + \boldsymbol{n} \otimes \boldsymbol{l} - \boldsymbol{m} \otimes \boldsymbol{\overline{m}} - \boldsymbol{\overline{m}} \otimes \boldsymbol{m} \, .  
\end{equation*}
In order to determine the connection in this formalism, we use the torsion-free Maurer--Cartan equation of structure
\begin{equation*} 
\textnormal{d}\boldsymbol{e}\indices{^{(a)}} = \gamma\indices{^{(a)}_{(b) (c)}} \boldsymbol{e}\indices{^{(b)}} \wedge \boldsymbol{e}\indices{^{(c)}} \, , 
\end{equation*}
in which the Ricci rotation coefficients $(\gamma\indices{^{(a)}_{(b) (c)}})$ are related to the connection via
\begin{equation*}
\gamma\indices{^{(a)}_{(b) (c)}} \boldsymbol{e}\indices{^{(c)}} = e\indices{_{\mu}^{(a)}} \, \textnormal{d}e\indices{^{\mu}_{(b)}} + e\indices{_{\mu}^{(a)}} \, e\indices{^{\nu}_{(b)}} \, \omega\indices{^{\mu}_{\nu}} \, .
\end{equation*}
Inserting the dual null basis, this equation becomes
\begin{align}
\textnormal{d}\boldsymbol{l} & = 2 \, \textnormal{Re}(\epsilon) \, \boldsymbol{n} \wedge \boldsymbol{l} - 2 \, \boldsymbol{n} \wedge \textnormal{Re}(\kappa \, \boldsymbol{\overline{m}}) - 2 \, \boldsymbol{l} \wedge \textnormal{Re}\bigl([\tau -\overline{\alpha} - \beta] \, \boldsymbol{\overline{m}}\bigr) + 2 \textnormal{i} \, \textnormal{Im}(\varrho) \, \boldsymbol{m} \wedge \boldsymbol{\overline{m}} \nonumber \\ \nonumber \\ 
\textnormal{d}\boldsymbol{n} & = 2 \, \textnormal{Re}(\gamma) \, \boldsymbol{n} \wedge \boldsymbol{l} - 2 \, \boldsymbol{n} \wedge \textnormal{Re}\bigl([\overline{\alpha} + \beta - \overline{\pi}] \, \boldsymbol{\overline{m}}\bigr) + 2 \, \boldsymbol{l} \wedge \textnormal{Re}(\overline{\nu} \,  \overline{\boldsymbol{m}}) + 2 \textnormal{i} \, \textnormal{Im}(\mu) \, \boldsymbol{m} \wedge \boldsymbol{\overline{m}} \label{NPMCE} \\ \nonumber \\
\textnormal{d}\boldsymbol{m} & = \overline{\textnormal{d}\boldsymbol{\overline{m}}} = (\overline{\pi} + \tau) \, \boldsymbol{n} \wedge \boldsymbol{l} + \bigl(2 \textnormal{i} \, \textnormal{Im}(\epsilon) - \varrho\bigr) \, \boldsymbol{n} \wedge \boldsymbol{m} - \sigma \, \boldsymbol{n} \wedge \boldsymbol{\overline{m}} + \bigl(\overline{\mu} + 2 \textnormal{i} \, \textnormal{Im}(\gamma)\bigr) \, \boldsymbol{l} \wedge \boldsymbol{m} \nonumber \\ 
& \hspace{1.4cm} + \overline{\lambda} \,  \boldsymbol{l} \wedge \boldsymbol{\overline{m}} - (\overline{\alpha} - \beta) \, \boldsymbol{m} \wedge \boldsymbol{\overline{m}} \, , \nonumber
\end{align}
where the so-called spin coefficients 
\begin{eqnarray*} 
\kappa = \gamma\indices{_{(2) (0) (0)}}  & \quad \varrho = \gamma\indices{_{(2) (0) (3)}} & \quad \epsilon = \tfrac{1}{2} \bigl(\gamma\indices{_{(1) (0) (0)}} + \gamma\indices{_{(2) (3) (0)}}\bigr) \nonumber \\
\sigma = \gamma\indices{_{(2) (0) (2)}} & \quad \mu = \gamma\indices{_{(1) (3) (2)}} & \quad \gamma = \tfrac{1}{2} \bigl(\gamma\indices{_{(1) (0) (1)}} + \gamma\indices{_{(2) (3) (1)}}\bigr) \nonumber \\
\lambda = \gamma\indices{_{(1) (3) (3)}} & \quad \tau = \gamma\indices{_{(2) (0) (1)}} & \quad \alpha = \tfrac{1}{2} \bigl(\gamma\indices{_{(1) (0) (3)}} + \gamma\indices{_{(2) (3) (3)}}\bigr) \\
\nu = \gamma\indices{_{(1) (3) (1)}} & \quad \pi = \gamma\indices{_{(1) (3) (0)}} & \quad \beta = \tfrac{1}{2} \bigl(\gamma\indices{_{(1) (0) (2)}} + \gamma\indices{_{(2) (3) (2)}}\bigr) \nonumber 
\end{eqnarray*}
constitute the connection representation in the Newman--Penrose formalism. Next, as tetrad frames may be subjected to local Lorentz transformations at every point on the manifold, we briefly address the particular tetrad transformations applied in Section \ref{Sec2}. These transformations are elements of the 2-parameter subgroup of local Lorentz transformations known as class III or spin-boost Lorentz transformations \cite{ChandraBook, BON}, which renormalize the real-valued Newman--Penrose vectors $\boldsymbol{l}$ and $\boldsymbol{n}$, leaving their directions unchanged, and rotate the complex-conjugate pair $\boldsymbol{m}$ and $\boldsymbol{\overline{m}}$ in the $(\boldsymbol{m}, \boldsymbol{\overline{m}})$-plane 
\begin{equation}\label{CLIIILLTT}
\boldsymbol{l} \mapsto \boldsymbol{l}' = \varsigma \, \boldsymbol{l} \, , \quad \boldsymbol{n} \mapsto \boldsymbol{n}' = \varsigma^{-1} \boldsymbol{n} \, , \quad \boldsymbol{m} \mapsto \boldsymbol{m}' = e^{\textnormal{i}\psi} \boldsymbol{m} \, , \quad \boldsymbol{\overline{m}} \mapsto \boldsymbol{\overline{m}}\hspace{0.03cm}' = e^{- \textnormal{i} \psi} \, \boldsymbol{\overline{m}} \, ,
\end{equation}
where the scale factor $\varsigma \in \mathbb{R} \backslash \{0\}$ and the angle $\psi \in \mathbb{R}$ are functions depending on the spacetime coordinates $(x^{\mu})$. There are various aspects of the Newman--Penrose formalism that are not discussed in this appendix, such as the different classes of local Lorentz transformations or the Weyl scalars and their algebraic classification. These can be found elsewhere in the literature. We refer the interested reader to \cite{BON, PRose}.

%-----------------------------------------------------------------------------------------------
\section{The General Relativistic Dirac Equation}\label{AB}
%-----------------------------------------------------------------------------------------------

\noindent The general relativistic, massive Dirac equation without an external potential is defined as \cite{Fock, Weyl} 
\begin{equation*}
(\gamma^{\mu} \nabla_{\mu} + \textnormal{i} m) \Psi(x^{\mu}) = \boldsymbol{0} \, ,
\end{equation*}
where $(\gamma^{\mu})$ are the general relativistic Dirac matrices, which satisfy the anticommutator relations 
\begin{equation*}
\{\gamma^{\mu}, \gamma^{\nu}\} = 2 g^{\mu \nu} \1_{\mathbb{C}^4} \, ,
\end{equation*}
$\boldsymbol{\nabla}$ is the metric connection on the spin bundle $S\mathfrak{M} = \mathfrak{M} \times \mathbb{C}^4$ on $\mathfrak{M}$ given by
\begin{equation*}
\boldsymbol{\nabla}_{\mu} = \partial_{\mu} + \frac{1}{8} \, \omega_{\mu \alpha \beta} \bigl[\gamma^{\alpha}, \gamma^{\beta} \hspace{0.02cm} \bigr] 
\end{equation*}
with $\boldsymbol{\omega}$ being the spin connection and $[ \, \cdot \,\, , \, \cdot \, ]$ the commutator, $\Psi$ is a Dirac $4$-spinor on sections $S_{\boldsymbol{x}}\mathfrak{M} \simeq \mathbb{C}^4$, $\boldsymbol{x} \in \mathfrak{M}$, of the spin bundle, and $m$ is the invariant fermion rest mass. Using the chiral $2$-spinor representation of the Dirac $4$-spinors and matrices \cite{FHM}
\begin{equation*}\label{ODS}
\Psi = \left(\begin{array}{c}
P^A \\
\overline{Q}_{B'} \\
\end{array}\right)
\quad \textnormal{and} \quad \gamma^{\mu} = \sqrt{2} \left(\begin{array}{cc}
\boldsymbol{0}_{\mathbb{C}^2} & \sigma\indices{^{\mu A B'}}\\
\sigma\indices{^{\mu}_{A B'}} & \boldsymbol{0}_{\mathbb{C}^2} \\
\end{array}\right) ,
\end{equation*}
in which $P^A$ and $\overline{Q}_{B'}$ denote $2$-component spinors, $(\sigma\indices{^{\mu}_{A B'}})$ are the Hermitian $(2 \times 2)$-Infeld--van der Waerden symbols, and $A \in \{1, 2\}$ as well as $B' \in \{1', 2'\}$, we obtain the following $2$-spinor form of the Dirac equation
\begin{equation}\label{SKDE}
\begin{split}
\nabla_{A B'} P^A + \frac{\textnormal{i} m}{\sqrt{2}} \, \overline{Q}_{B'} & = \boldsymbol{0} \\ 
\nabla_{A B'} Q^A + \frac{\textnormal{i} m}{\sqrt{2}} \, \overline{P}_{B'} & = \boldsymbol{0} \, ,
\end{split}
\end{equation}
where $\nabla_{A B'} = \sigma\indices{^{\mu}_{A B'}} \nabla_{\mu}$. In this notation, quantities with primed indices are subjected to the complex conjugates of the transformations that are applied to those labeled with unprimed indices. In order to express these equations in the framework of the Newman--Penrose formalism, we now introduce the local dyad basis $(\zeta_{(k)})$ and its dual $(\zeta^{(k)})$, $k \in \{1, 2\}$, for the space of Dirac 2-spinors. The associated local spinor components $\mathscr{Y}^{(k)}$ and $\mathscr{Y}_{(k)}$ are related to the previous $2$-spinor components $\mathscr{Y}^{A}, \mathscr{Y}_{A} \in \mathbb{C}^2$ via the $(2 \times 2)$-matrix $\zeta\indices{^{(k)}_{A}}$ and its inverse $\zeta\indices{_{(k)}^{A}}$ in a similar vein as in the tetrad formalism by 
\begin{equation*}
\mathscr{Y}^{(k)} = \zeta\indices{^{(k)}_{A}} \mathscr{Y}^{A} \quad \textnormal{and} \quad \mathscr{Y}_{(k)} = \zeta\indices{_{(k)}^{A}} \, \mathscr{Y}_{A} \, .
\end{equation*}
In this representation, the metric connection reads 
\begin{equation}\label{dyadicmc}
\nabla\indices{_{(k) (l')}} \mathscr{Y}^{(m)} = \zeta\indices{^{A}_{(k)}} \overline{\zeta}\indices{^{B'}_{(l')}} \, \zeta\indices{^{(m)}_{C}} \nabla_{A B'} \, \mathscr{Y}^C = \partial_{(k) (l')} \mathscr{Y}^{(m)} + \Gamma\indices{^{(m)}_{(n) (k) (l')}} \mathscr{Y}^{(n)} \,,
\end{equation}
where $\partial_{(k) (l')} = \sigma\indices{^{\mu}_{(k) (l')}} \partial_{\mu}$ and 
\begin{equation}\label{dyadicSC}
\Gamma\indices{^{(m)}_{(n) (k) (l')}} = \Gamma\indices{^{(m) (o')}_{(n) (o') (k) (l')}} = \sqrt{2} \, 
\epsilon\indices{^{(m) (q)}} \, \epsilon\indices{^{(o') (p')}} \, \sigma\indices{^{\mu}_{(q) (p')}} \, \sigma\indices{^{\nu}_{(n) (o')}} \, \sigma\indices{^{\lambda}_{(k) (l')}} \, e\indices{_{\mu}^{(a)}} \, e\indices{_{\nu}^{(b)}} \, e\indices{_{\lambda}^{(c)}} \, \gamma\indices{_{(a) (b) (c)}} \, . 
\end{equation}
We remark that the $2$-dimensional Levi--Civita symbol $\boldsymbol{\epsilon}$ acts as skew metric on $\mathbb{C}^2$. Moreover, the Infeld--van der Waerden symbols take the form
\begin{equation}\label{IVDWS}
\sigma\indices{^{\mu}_{(k) (l')}} = \left(\begin{array}{cc}
l^{\mu} & m^{\mu} \\
\overline{m}^{\, \mu} & n^{\mu} \\
\end{array}\right).
\end{equation}
Employing (\ref{dyadicmc})-(\ref{IVDWS}) in (\ref{SKDE}) and using the notation $\mathscr{F}_1 := P^{(1)}, \mathscr{F}_2 := P^{(2)}, \mathscr{G}_1 := \overline{Q}^{\, (2')},$ as well as $\mathscr{G}_2 := - \overline{Q}^{\, (1')}$, the general relativistic Dirac equation in the Newman--Penrose formalism becomes
\begin{align} \label{NPCSTDE}
\begin{split}
(l^{\mu} \partial_{\mu} + \varepsilon - \varrho) \mathscr{F}_1 + (\overline{m}^{\mu} \partial_{\mu} + \pi - \alpha) \mathscr{F}_2 & = \frac{\textnormal{i} m}{\sqrt{2}} \, \mathscr{G}_1 \\ 
(n^{\mu} \partial_{\mu} + \mu - \gamma) \mathscr{F}_2 + (m^{\mu} \partial_{\mu} + \beta - \tau) \mathscr{F}_1 & = \frac{\textnormal{i} m}{\sqrt{2}} \, \mathscr{G}_2 \\ 
(l^{\mu} \partial_{\mu} + \overline{\varepsilon} - \overline{\varrho}) \mathscr{G}_2 - (m^{\mu} \partial_{\mu} + \overline{\pi} - \overline{\alpha}) \mathscr{G}_1 & = \frac{\textnormal{i} m}{\sqrt{2}} \, \mathscr{F}_2 \\ 
(n^{\mu} \partial_{\mu} + \overline{\mu} - \overline{\gamma}) \mathscr{G}_1 - (\overline{m}^{\mu} \partial_{\mu} + \overline{\beta} - \overline{\tau}) \mathscr{G}_2 & = \frac{\textnormal{i} m}{\sqrt{2}} \, \mathscr{F}_1 \, . 
\end{split}
\end{align}

\end{appendix}


\vspace{0.5cm}

\begin{thebibliography}{99}

\bibitem{Bat}
D.~Batic, ``Scattering for massive Dirac fields on the Kerr metric,'' Journal of Mathematical Physics {\bf 48}, 022502 (2007).

\bibitem{BL}
R.\ H.~Boyer and R.\ W.~Lindquist, ``Maximal analytic extension of the Kerr metric,'' Journal of Mathematical Physics {\bf 8}, 265 (1967).
  
\bibitem{BriWhee}
D.\ R.~Brill and J.\ A.~Wheeler, ``Interaction of neutrinos and gravitational fields,'' Review of Modern Physics {\bf 29}, 465 (1957).  
  
\bibitem{Carter}
B.~Carter, ``Black hole equilibrium states,'' In: C.~DeWitt and B.\ S.~DeWitt, ``Black holes/Les astres occlus,'' Gordon and Breach (1973).
   
\bibitem{ChaMuk}
S.\ K.~Chakrabarti and B.~Mukhopadhyay, ``Scattering of Dirac waves off Kerr black holes,'' Monthly Notices of the Royal Astronomical Society {\bf 317}, 979 (2000).  

\bibitem{Chandra}
S.~Chandrasekhar, ``The solution of Dirac's equation in Kerr geometry,'' Proceedings of the Royal Society London {\bf 349}, 571 (1976).
  
\bibitem{ChaDet}
S.~Chandrasekhar and S.~Detweiler, ``On the reflexion and transmission of neutrino waves by a Kerr black hole,'' Proceedings of the Royal Society London {\bf 352}, 325 (1977).  

\bibitem{ChandraBook} 
S.~Chandrasekhar, ``The mathematical theory of black holes,'' Oxford University Press (1983). 

\bibitem{CoddLevi}
E.\ A.~Coddington and N.~Levinson, ``Theory of ordinary differential equations,'' Tata McGraw--Hill Publishing Company Limited (1972).

\bibitem{Daude}
T.~Daud{\'e}, ``Propagation estimates for Dirac operators and application to scattering theory,'' Annales de l'institut Fourier {\bf 54}, 2021 (2004).

\bibitem{DolGair}
S.\ R.~Dolan and J.\ R.~Gair, ``The massive Dirac field on a rotating black hole spacetime: angular solutions,'' Classical and Quantum Gravity {\bf 26}, 175020 (2009).
    
\bibitem{Dolan3}
S.\ R.~Dolan and  D.\ Dempsey, ``Bound states of the Dirac equation on Kerr spacetime,'' Classical and Quantum Gravity {\bf 32}, 184001 (2015).

\bibitem{Doran}
C.~Doran, ``New form of the Kerr solution,'' Physical Review D {\bf 61}, 067503 (2000).	
	
\bibitem{Edd}
A.\ S.~Eddington, ``A comparison of Whitehead's and Einstein's formulae,'' Nature {\bf 113}, 192 (1924).

\bibitem{Fink}
D.~Finkelstein, ``Past-future asymmetry of the gravitational field of a point particle,'' Physical Review {\bf 110}, 965 (1958).	
	
\bibitem{FKSM1}
F.~Finster, N.~Kamran, J.~Smoller, and S.\ T.~Yau, ``Non-existence of time-periodic solutions of the Dirac equation in an axisymmetric black hole geometry,'' Communications on Pure and Applied Mathematics {\bf 53}, 902 (2000).

\bibitem{FKSM2}
F.~Finster, N.~Kamran, J.~Smoller, and S.\ T.~Yau, ``Decay rates and probability estimates for massive Dirac particles in the Kerr--Newman black hole geometry,'' Communications in Mathematical Physics {\bf 230}, 201 (2002).
  
\bibitem{FKSM3}
F.~Finster, N.~Kamran, J.~Smoller, and S.\ T.~Yau, ``The long-time dynamics of Dirac particles in the Kerr--Newman black hole geometry,'' Advances in Theoretical and Mathematical Physics {\bf 7}, 25 (2003). 

\bibitem{RF} 
F.~Finster and C.~R\"oken, ``An integral spectral representation of the massive Dirac propagator in the Kerr geometry in Eddington--Finkelstein-type coordinates,'' Advances in Theoretical and Mathematical Physics {\bf 22}, 47 (2018).      
   
\bibitem{Fock}
V.~Fock, ``Geometrisierung der Diracschen Theorie des Elektrons,'' Zeitschrift f\"ur Physik {\bf 57}, 261 (1929).
    
\bibitem{FHM}
J.\ A.\ H.~Futterman, F.\ A.~Handler, and R.\ A.~Matzner, ``Scattering from black holes,'' Cambridge University Press (1988).
  
\bibitem{GMNRS}
J.\ N.~Goldberg, A.\ J.~Macfarlane, E.\ T.~Newman, F.~Rohrlich, and E.\ C.\ G.~Sudarshan, ``Spin-s spherical harmonics and $\eth$,'' Journal of Mathematical Physics {\bf 8}, 2155 (1967).	

\bibitem{DH}
D.~H\"afner, ``Creation of fermions by rotating charged black-holes,'' arXiv:math/0612501 [math.AP] (2006).

\bibitem{HaNic}
D.~H\"afner and J.--P.~Nicolas, ``Scattering of massless Dirac fields by a Kerr black hole,'' Reviews in Mathematical Physics {\bf 16}, 29 (2004). 
  
\bibitem{Kerr}
R.\ P.~Kerr, ``Gravitational field of a spinning mass as an example of algebraically special metrics,'' Physical Review Letters {\bf 11}, 237 (1963).  
      
\bibitem{Kinn}
W.~Kinnersley, ``Type D vacuum metrics,'' Journal of Mathematical Physics {\bf 10}, 1195 (1969).  

\bibitem{Kruskal}
M.\ D.~Kruskal, ``Maximal extension of Schwarzschild metric,'' Physical Review {\bf 119}, 1743 (1960).

\bibitem{NP}
E.\ T.~Newman and R.~Penrose, ``An approach to gravitational radiation by a method of spin coefficients,'' Journal of Mathematical Physics {\bf 3}, 566 (1962).

\bibitem{Nic}
J.--P.~Nicolas, ``Dirac fields on asymptotically simple space-times,'' Jean Leray '99 Conference Proceedings {\bf 24}, 205 (2003).

\bibitem{BON}
B.~O'Neill, ``The geometry of Kerr black holes,'' Dover Publications (2014).

\bibitem{Page}
D.\ N.~Page, ``Dirac equation around a charged, rotating black hole,'' Physical Review D {\bf 14}, 1509 (1976). 
  
\bibitem{PRose}
R.~Penrose, ``A spinor approach to general relativity,'' Annals of Physics {\bf 10}, 171 (1960). 
  
\bibitem{PrTe}
W.\ H.~Press and S.\ A.~Teukolsky, ``Perturbations of a rotating black hole. II. Dynamical stability of the Kerr metric,'' Astrophysical Journal {\bf 185}, 649 (1973).   

\bibitem{CR}
C.~R\"oken, ``Kerr isolated horizons in Ashtekar and Ashtekar--Barbero connection variables,'' General Relativity and Gravitation {\bf 49}, id.\ 114 (2017).

\bibitem{Teuk}
S.\ A.~Teukolsky, ``Perturbations of a rotating black hole. I. Fundamental equations for gravitational, electromagnetic, and neutrino-field perturbations,'' Astrophysical Journal {\bf 185}, 635 (1973).

\bibitem{Unruh}
W.\ G.~Unruh, ``Separability of the neutrino equations in a Kerr background,'' Physical Review Letters {\bf 31}, 1265 (1973).

\bibitem{Weyl}
H.~Weyl, ``Elektron und Gravitation,'' Zeitschrift f\"ur Physik {\bf 56}, 330 (1929).

\end{thebibliography}
\end{document}